\documentclass[aps,11pt,nofootinbib,preprintnumbers]{revtex4}
\usepackage[dvipdfmx]{graphicx}
\usepackage{amssymb,amsfonts,amsmath,bm,color,multirow,cases,empheq,hyperref}
\usepackage{mathrsfs}
\usepackage{here}

\usepackage{enumerate}
\usepackage{ulem}
\usepackage{afterpage}
\hypersetup{%
 setpagesize=false,
 bookmarksnumbered=true,%
 bookmarksopen=true,%
 colorlinks=true,%
 linkcolor=blue,%
 citecolor=red}
\if0

\usepackage{amsmath,amssymb}
\usepackage{graphicx,enumerate,color}

\fi

\newcommand{\fr}[1]{\frac{1}{#1}}

\newcommand{\ord}[1]{{\mathcal O}(#1)}

\newcommand{\cR}{{\mathcal R}}
\newcommand{\cL}{{\mathcal L}}

\newcommand{\nonum}{\nonumber\\ }

\newcommand{\cout}[1]{}
\preprint{TTI-MATHPHYS-14}
\begin{document}

 \title{Einstein-Gauss-Bonnet black strings at large $\alpha$}
\author{Ryotaku Suzuki}
\email{sryotaku@toyota-ti.ac.jp}
\author{Shinya Tomizawa}
\email{tomizawa@toyota-ti.ac.jp}
\affiliation{Mathematical Physics Laboratory Toyota Technological Institute\\
Hisakata 2-12-1, Nagoya 468-8511, Japan}
\date{\today}

\begin{abstract}

The simplest black string in higher-dimensional general relativity (GR) is perhaps the direct product of a Schwarzschild spacetime and a flat spatial direction. 
However, it is known that the Einstein-Gauss-Bonnet theory does not allow such a trivial and simple solution. 
We propose a novel analytic technique, which assumes that the Gauss-Bonnet (GB) term becomes dominant over the Einstein-Hilbert (EH) term.
Assuming the dimensionless coupling constant $\alpha$ normalized by the horizon scale is large enough,
we find that the spacetime is separated into the GB region and GR region, which are matched via the transition region where the GB and EH terms are comparable.
Using this {\it large $\alpha$ approximation}, we indeed construct new analytic solutions of black strings, from which we analytically compute various physical quantities of black strings at large $\alpha$. 
Moreover, we confirm that all these analytic results are consistent with the numerical calculation. 
We also discuss the possible extension to general Einstein-Lovelock black holes.

\end{abstract}

\vskip 2cm
\pagestyle{empty}
\small

%\addtocontents{toc}{\protect\setcounter{tocdepth}{2}}
%{
%	\hypersetup{linkcolor=black,linktoc=all}
%	\tableofcontents
%}
%\normalsize
%\newpage
\pagestyle{plain}
\setcounter{page}{1}

\maketitle

\section{Introduction}

For two decades, black holes in dimensions more than four have attracted many physicists from a point of view of  scientific and applied researches,  for instance, by the microscopic derivation of
Bekenstein-Hawking entropy~\cite{Strominger:1996sh},  the realistic production of black holes at accelerators in the
scenario of large extra dimensions~\cite{Argyres:1998qn}, and  AdS/CFT correspondence~\cite{Maldacena:1997re}.
The recent developments by many researchers show  the richness of such solutions and the complexity of the dynamics~\cite{Emparan:2008eg}. 
In particular, a  black string is one of the simplest non-spherical  black objects which can be constructed by simply adding a flat spatial direction to the Schwarzschild-Tangherlini black hole~\cite{Tangherlini:1963bw}.
For simplicity, it has been a major testing ground for the black hole dynamics proper to higher dimensional black holes such as the Gregory-Laflamme instability~\cite{Gregory:1993vy,Gregory:1994bj}.

\medskip
So far, higher-dimensional black holes are studied mostly in Einstein's theory of general relativity (GR). However, GR is not a unique theory of gravity in higher dimensions, but it can also include higher curvature corrections, %which are string theory inspired corrections in the ultraviolet scale~\cite{Garraffo:2008hu}.
which are inspired from the string theory in the ultraviolet scale~\cite{Garraffo:2008hu}.
Einstein-Lovelock theories are one of such generalization of GR to the higher curvature theories whose equations of motion are  second order differential equations. 
It is of great interest in theoretical physics as it describes a wide class of models. 
In Einstein-Lovelock theories,  the only found exact solutions of black holes are static and spherically symmetric black hole solutions due to the lack of analytical methods.
Moreover, even black string solutions cannot be obtained by the simple addition of a flat spatial direction~\cite{Barcelo:2002wz}. However, it is known that black strings admit such simple construction in the pure Lovelock theories that only have a single Lovelock term without Einstein-Hilbert (EH) term in the action~\cite{Kastor:2006vw,Giribet:2006ec}.
The inclusion of the negative cosmological constant and supporting scalar or $p$-form fields also admits another types of homogeneous Einstein-Lovelock black branes~\cite{Cisterna:2018mww,Cisterna:2020kde,Canfora:2021ttl,Cisterna:2021ckn}, which are hinted by the construction of homogeneous AdS black branes supported by scalar fields~\cite{Cisterna:2017qrb}.

\medskip
In this article, we focus on the $d\ (>4)$-dimensional Einstein-Lovelock theory with only quadratic curvature corrections, i.e., the $d$-dimensional Einstein-Gauss-Bonnet (EGB) theory, whose action is given by
\begin{align}
 S = \fr{16 \pi G} \int d^dx \sqrt{-g} (R + \alpha_{\rm GB} \cL_{\rm GB}),
 \label{eq:EGBaction}
\end{align}
where the Gauss-Bonnet (GB) term is written as
\begin{align}
 \cL_{\rm GB} := R^2 - 4 R_{\mu\nu} R^{\mu\nu} + R_{\mu\nu\rho\sigma}R^{\mu\nu\rho\sigma}.
  \label{eq:GBterm}
\end{align}
In the EGB theory, a fundamental parameter of the theory is the coupling constant $\alpha_{\rm GB}$. 
Hence it might be natural to study  to construct black strings for small coupling constant~\cite{Kobayashi:2004hq,Brihaye:2010me,Kleihaus:2012qz}. 
This limit is also used for the construction of spinning black holes in the five-dimentional EGB theory~\cite{Ma:2020xwi}. 
Another controllable parameter is the dimension $d$, which leads to the large $d$ limit~\cite{Emparan:2013moa,Emparan:2020inr} that is also useful in the construction of EGB black strings~\cite{Chen:2017rxa,Chen:2018vbv} and EGB rotating black holes~\cite{Suzuki:2022apk}.
Besides these limiting solutions, EGB black string solutions were numerically obtained as well~\cite{Kobayashi:2004hq,Brihaye:2010me,Kleihaus:2012qz}.

\medskip
Now let us consider another new possibility of the parameter limit, in which the GB term becomes dominant over the EH term around black holes or certain compact objects
\begin{align}
  R \ll \alpha_{\rm GB} \cL_{\rm GB}. \label{eq:GB-dominant}
\end{align}
This leads us to an interesting approximation,
 which we call {\it the large $\alpha$ approximation}, where $\alpha$ is the dimensionless coupling constant normalized by the spacetime curvature around the system $R\sim \cR$
\footnote{We only consider the positive $\alpha_{\rm GB}$, since the coupling constant is bound below in the negative case.}
\begin{align}
 \alpha \sim \alpha_{\rm GB} \cR \gg 1.
\end{align}

For black holes of the radius $r_0$, it should simply be $\cR \sim 1/r_0^2$,
and hence, one can also interpret this as the large curvature approximation or small black hole approximation when $\alpha_{\rm GB}$ is fixed.
At the limit $\alpha\to \infty$, it is natural to expect that spacetime geometry approaches that of the pure GB theory, in which black strings restore a simple construction with a flat spatial direction~\cite{Kastor:2006vw,Giribet:2006ec}.
Indeed, static EGB black holes approach static pure GB black holes at large $\alpha$~\cite{Giacomini:2015dwa}.
 The big difference from the pure GB theory is that arbitrarily large but finite $\alpha$ causes the breakdown of the assumption~(\ref{eq:GB-dominant}) near the flat region where the spacetime becomes almost GR
\begin{align}
  R \gg \alpha_{\rm GB} {\cal L}_{\rm GB}.
\end{align}
In our previous study on the stable bound orbits around the spherical EGB black hole~\cite{Suzuki:2022snr}, we found that
these two regions,
which from now on we call {\it the GB region} and {\it the GR region},
can be matched by an intermediate region of the {\it transition region}. 
In this article, we use the large $\alpha$ approximation to construct EGB black strings. 
More precisely, we solve the EGB equations in the GB, transition and GR region analytically, and then match them to obtain the entire geometry.
The physical quantities are also derived up to the leading order and compared with the numerical calculation in $d=6,\dots,10$. 
To see the validity of  the large $\alpha$ approximation, we confirm whether the analytical result fits well with the numerical result in the large $\alpha$ regime.

\medskip
The rest of this article  is organized as follows. We first explain the setup in section~\ref{sec:setup}.
In section~\ref{sec:largealphaBH}, we revisit EGB black holes at large $\alpha$. 
Then, the GB region of EGB black strings is solved at large $\alpha$ in section~\ref{sec:largealpha}. In section~\ref{sec:transition}, we study the matching in the transition region to connect the metric in the GB region and asymptotically flat background, and then obtain the expression for the physical quantities. These analytic results are compared with the numerical calculation in section~\ref{sec:numeric}. Finally, we discuss the possible extension to Einstein-Lovelock black holes in section~\ref{sec:lovelockbh}. The results are summarized in section~\ref{sec:summary}.

\section{Setup}\label{sec:setup}

From the action~(\ref{eq:EGBaction}), the EGB equation can be derived as
\begin{align}
 R_{\mu\nu} - \fr{2} R g_{\mu\nu} + \alpha_{\rm GB} H_{\mu\nu} =0\label{eq:EGBeq}
\end{align}
where $H_{\mu\nu}$ is the Lanzcoz tensor given by
\begin{align}
 H_{\mu\nu} = 2 R R_{\mu\nu} - 4 R_{\mu \alpha} R^\alpha{}_\nu - 4 R_{\mu \alpha \nu \beta} R^{\alpha\beta} + 2 R_{\mu \alpha\beta\gamma} R_\nu{}^{\alpha\beta\gamma}-\fr{2} \cL_{\rm GB} g_{\mu\nu}.
\end{align}
Let us consider a $d=n+4$ uniform black string under the ansatz following the convention in ref.~\cite{Brihaye:2010me} 
\begin{align}
ds^2 = - b(r)dt^2 + \frac{dr^2}{f(r)}+a(r) dz^2 + r^2 d\Omega_{n+1}^2.\label{eq:ansatz-string}
\end{align}
We assume that the metric asymptotes to the Kaluza-Klein background $M_{n+3} \times S^1$, i.e., the direct product of a $(n+3)$-dimensional Minkowski spacetime and a flat compact direction,  which can be written at $r\to \infty$ as
\begin{align}
  a \to 1,\quad b\to 1,\quad f\to 1,
\end{align}
with $z$-direction identified by $z\sim z+ L$.
The asymptotic behavior to this background is solved as
\begin{align}
 b = 1 -\frac{c_t}{r^n},\quad a = 1 + \frac{c_z}{r^n} ,\quad f = 1 - \frac{c_t-c_z}{r^n}. \label{eq:GR-asym}
\end{align}
This enables us to determine the mass $M$ and tension $\cal T$
by using the Arnowitt-Deser-Misner (ADM) formula in~\cite{Harmark:2004ch}, 
\begin{align}
M = \frac{\Omega_{n+1}L}{16\pi G}((n+1)c_t-c_z),\quad 
{\cal T} = \frac{\Omega_{n+1}}{16 \pi G}(c_t-(n+1)c_z)\label{eq:adm-mass-tau}.
\end{align}
The so-called relative tension, or relative binding energy, is also computed as
\begin{align}
 N = \frac{{\cal T} L}{M}=\frac{c_t-(n+1)c_z}{(n+1)c_t-c_z}.
\end{align}
The Hawking temperature $T_{\rm H}$ is defined by the surface gravity 
\begin{align}
   T_{\rm H} = \frac{\kappa}{2\pi}= \fr{4\pi r_0} \sqrt{f'(r_0)a'(r_0)},
\end{align}
where $r=r_0$ is the horizon radius.
Moreover, the black hole entropy  can be derived by the Iyer-Wald formula~\cite{Wald:1993nt,Iyer:1994ys} as
\begin{align}
S = \fr{4G} \int_H \left (1+2\alpha_{\rm GB} \cR\right)dS = \frac{\Omega_{n+1}}{4G}  r_0^{n+1}\left(1+ \frac{2n(n+1)\alpha_{\rm GB}}{r_0^2}\right) \sqrt{a(r_0)}
\end{align}
where $\cR$ is the spatial curvature on the horizon.

\section{EGB black holes at large $\alpha$}\label{sec:largealphaBH}
\begin{figure}[t]
\centering{
\includegraphics[width=14cm]{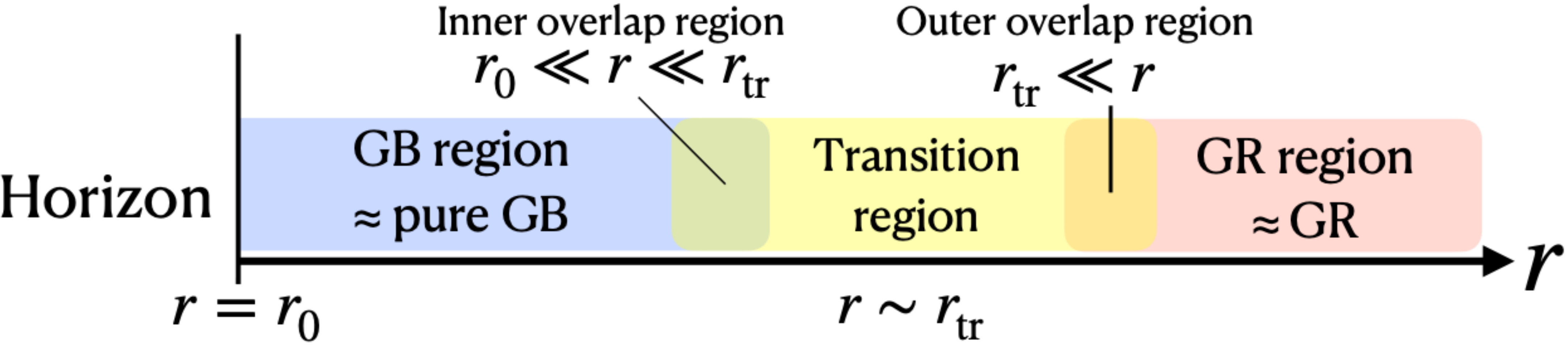}
\caption{Two separate regions in EGB black holes at large $\alpha$.\label{fig:sch-EGB}}
}
\end{figure}

In our previous study on stable bound orbits around the spherical EGB black holes~\cite{Boulware:1985wk}, we have found that the large $\alpha$ approximation provides us a useful analytic approach,  which simplifies the analysis of the geodesic motion~\cite{Suzuki:2022snr}. Before the black string analysis,
we revisit the large $\alpha$ limit for asymptotically flat, static and spherically symmetric EGB black holes in the $(n+3)$-dimension. 
The metric of  the EGB black hole spacetime is given by
\begin{align}
  ds^2 = -F(r)dt^2 + \frac{dr^2}{F(r)} + r^2 d\Omega^2_{n+1},\label{eq:EGBBH-metric}
\end{align}
with
\begin{align}
 F(r) = 1+\frac{r^2}{2\alpha r_0^2} \left(1-\sqrt{1+\frac{4\alpha(\alpha+1)r_0^{n+2}}{r^{n+2}}}\right),
 \label{eq:EGBBH-Fr}
\end{align}
where
the horizon is at $r=r_0$, and  we have introduced the dimensionless coupling constant by
\begin{align}
 \alpha := \frac{n(n-1)\alpha_{\rm GB}}{r_0^2}.\label{eq:def-alpha}
\end{align}
In Ref.~\cite{Giacomini:2015dwa}, it is noticed that this solution is endowed with the following two regimes.
For the large $\alpha$ with fixed $r$, the solution approaches a black hole spacetime in the pure GB theory
\begin{align}
 F(r) \simeq 1 - \left(\frac{r_0}{r}\right)^\frac{n-2}{2}.\label{eq:Fr-pureGB}
\end{align}
On the other hand, the limit $r\to\infty$ with fixed $\alpha$ recovers the asymptotic behavior in $d=n+3$ GR
\begin{align}
F(r) \simeq 1 - \frac{(\alpha+1) r_0^{n}}{r^n}.
\end{align}
%As pointed out in Ref.~\cite{Suzuki:2022snr},
In Ref.~\cite{Suzuki:2022snr}, it is further pointed out that two regions, where we call the GB region and the GR region,  are separated by the transition scale 
\begin{align}
r_{\rm tr} = r_0\, \alpha^\frac{2}{n+2}, \label{eq:trans_scale}
\end{align}
which can be seen from Fig~\ref{fig:sch-EGB}. Around this scale $r\sim r_{\rm tr}$ a transition region is formed, which has sufficient overlaps with other two regions to allow the matched asymptotic expansion.
For $n=2$ ($d=5$), the pure GB theory becomes kinematic as the three dimensional GR, and there is no black hole solution without negative cosmological constant~
\cite{Dadhich:2012cv,Kastor:2012se}.
Nevertheless, eq.~(\ref{eq:EGBBH-Fr}) still admits a horizon in the large $\alpha$ approximation for $n=2$
\begin{align}
 F(r) \simeq \frac{r^2-r_0^2}{2\alpha  r_0^2}.\label{eq:BH-n2-limit}
\end{align}
Unlike $n>2$ cases, we cannot take the limit $\alpha \to \infty$ in $F(r)$.
Taking into account such difference, we consider $n>2$ cases and $n=2$ separately in the
following analysis.

\section{EGB black strings in GB region}\label{sec:largealpha}

First, we consider the GB region.
Although a simple direct product of the $(n+3)$-dimensional black hole metric~(\ref{eq:EGBBH-metric}) with a one-dimensional flat metric is not a solution to the $(n+4)$-dimensional EGB equation~(\ref{eq:EGBeq}),  
we expect that the spatial section of a $(n+4)$-dimensional black string~(\ref{eq:ansatz-string}) has the same separation of scales as the $(n+3)$-dimensional black hole  in the transverse direction. 
Note that we  consider only $n\geq 2$ ($d\geq6$) since for $n=1\ (d=5)$ $\alpha_{\rm GB}$ is bounded above by the horizon scale~\cite{Kobayashi:2004hq}\footnote{
Interestingly, ref.~\cite{Suranyi:2008wc} investigates black strings close to the maximum $\alpha_{\rm GB,max}$ in five dimension.}, and hence our approach cannot be used.
We introduce the dimensionless constant $\alpha$ by eq.~(\ref{eq:def-alpha})  in which  the black hole radius is replaced with the black string radius. 
In the following, as mentioned in the previous section, we study the $n>2$ cases and $n=2$ case separately by using the $1/\alpha$ expansion.

\subsection{$n>2$ cases}
For $n>2$, the black string solution for the pure GB theory is easily found as~\cite{Kastor:2006vw}
\begin{align}
 b(r) = f(r) = 1 -\left( \frac{r_0}{r}\right)^\frac{n-2}{2}  ,\quad  a(r) = 1, \label{eq:pureGBgtr2}
\end{align}
where we set the horizon at $r=r_0$. 
Then, we consider the $1/\alpha$-correction
\begin{align}
  b =b_H \left( 1- \left(\frac{r_0}{r}\right)^\frac{n-2}{2}+\fr{\alpha}b_1\right),\quad f = 1- \left(\frac{r_0}{r}\right)^\frac{n-2}{2}+\fr{\alpha}f_1,\quad a = a_H \left(1+\fr{\alpha}a_1\right)\label{eq:largealpha-expand-ngtr2}
\end{align}
where $b_H$ and $a_H$ are the constant scales of $t$ and $z$, which are determined by the asymptotic behavior.
From the combination of some equations of motion (see appendix~\ref{app:exp-eqn}), the equations for $a_1$, $b_1$ and $f_1$ are written as, respectively, 
\begin{align}
 \frac{d}{dr}\left[ r\left(\left(\frac{r}{r_0}\right)^\frac{n-2}{2}-1\right) \frac{d}{dr} a_1\right] = \frac{n+4}{2n(n+1)r_0}\left(\frac{r}{r_0}\right)^\frac{n}{2},\label{eq:largeA-corr-eq-a1} 
\end{align}
\begin{align}
\partial_r \left( \frac{b_1}{1 - (r/r_0)^{-\frac{n-2}{2}}}\right)
&= \frac{2n-(n+2)(r/r_0)^{-\frac{n-2}{2}}}{4 (n-1)  \left(1-(r/r_0)^{-\frac{n-2}{2}}\right)}a_1'\notag \\
   &-\frac{(n-2) f_1 }{2 r \left( 1-(r/r_0)^{-\frac{n-2}{2}}\right)^2}
   +\frac{(n+2) r}{4 (n-1) n r_0^2 \left(1-(r/r_0)^{-\frac{n-2}{2}}\right)},\label{eq:largeA-corr-eq-b1} 
\end{align}

\begin{align}
 f_1 = \frac{(n-2)(n+3)(r/r_0)^2}{2n(n+2)(n^2-1)} + \frac{C_1}{r^{\frac{n-2}{2}}}+\frac{n-2}{4(n-1)}\left(\frac{r_0}{r}\right)^\frac{n-2}{2}a_1,\label{eq:largeA-corr-eq-f1-int} 
\end{align}
where $C_1$ is an integration constant.
The regular solution of eq.~(\ref{eq:largeA-corr-eq-a1}) becomes
\begin{align}
 a_1 = \frac{2(n+4)}{(n-2)n(n+1)(n+2)} {\sf F}_n\left((r_0/r)^\frac{n-2}{2}\right),\label{eq:largeA-corr-sol-a1}
\end{align}
where we defined
\begin{align}
{\sf F}_n(x) := \int^1_x \frac{y^\frac{n+2}{2-n}-1}{1-y}dy.\label{eq:largeA-def_Fn}
\end{align}
Then, $f_1$ and $b_1$ are solved as,  respectively, 
\begin{align}
&  b_1 =\frac{(r/r_0)^2-1}{2n(n-1)}-\frac{8+5n}{2n(n-2)(n^2-1)}\left(1-\left(\frac{r_0}{r}\right)^{\frac{n-2}{2}}\right) - \frac{(n+4)(4-(n+2)(r_0/r)^\frac{n-2}{2}){\sf F}_n\left((r_0/r)^\frac{n-2}{2}\right)}{2n(n^2-1)(n^2-4)},\label{eq:largeA-corr-sol-b1}\\
& f_1 =  \frac{(n-2)(n+3)r^2}{2n(n+2)(n^2-1)r_0^2} - \frac{(n-2)(n-3)-(n+4){\sf F}_n\left((r_0/r)^\frac{n-2}{2}\right)}{2n(n^2-1)(n+2)}\left(\frac{r_0}{r}\right)^{\frac{n-2}{2}},\label{eq:largeA-corr-sol-f1}
\end{align}
where the integration constants are determined by $b_1(r_0)=b_1'(r_0)=f_1(r_0)=a_1(r_0)=0$. All the scaling degrees of freedom for $t\to \lambda_1 t$ and $z\to \lambda_2 z$ are absorbed into $a_H$ and $b_H$.

\subsection{$n=2$ case}
In the $n=2$ case, the black hole result~(\ref{eq:BH-n2-limit}) suggests that the metric function should scale as $f = \ord{\alpha^{-1}}$.
This rescaling actually leads to the leading order solution at large $\alpha$
\begin{align}
 a = a_H \frac{r^2}{r_0^2}  ,\quad f = \frac{1}{6\alpha} \left(\frac{r^2}{r_0^2}-\frac{r_0}{r}\right),\quad b = b_H \left(\frac{r^2}{r_0^2}-\frac{r_0}{r}\right),\label{eq:GBsol-n2-0}
\end{align}
where $a_H$ and $b_H$ are again the scaling constant of $z$ and $t$.
Clearly, this leading order metric cannot have asymptotically flat region at $r\to\infty$, but rather tends to be a warped product of $AdS_3$ and $S^{3}$,
\begin{align}
 ds^2\simeq 6 \alpha r_0^2\left[\frac{dr^2}{r^2} + \frac{r^2}{r_0^2} \left(- d\bar{t}^2+d\bar{z}^2\right)\right]+ r^2 d\Omega^2_{3},
 \quad \bar{t} := \sqrt{\frac{b_H}{6\alpha}} \frac{t}{r_0},\quad \bar{z} := \sqrt{\frac{a_H}{6\alpha}} \frac{t}{r_0}.
\end{align}
The $1/\alpha$ correction
\begin{align}
 a = a_H \left(\frac{r^2}{r_0^2}+\frac{a_1}{\alpha}\right)  ,\quad f = \frac{1}{6\alpha} \left(\frac{r^2}{r_0^2}-\frac{r_0}{r}\right)+\frac{ f_1}{\alpha^{2}},\quad b = b_H \left(\frac{r^2}{r_0^2}-\frac{r_0}{r}+\frac{b_1}{\alpha}\right),
\end{align}
is easily found as
\begin{subequations}\label{eq:GBsol-n2-1}
\begin{align}
&a_1 = \frac{r}{2r_0} \,
   _2F_1\left(\frac{1}{3},1,\frac{4}{3};\frac{r_0^3}{r^3}\right)
   -\frac{r(r^3+r_0^3) }{6r_0^4}+\frac{r^2}{6r_0^2}  \log \left(1-\frac{r_0^3}{r^{3}}\right),\\
&b_1 = \frac{ (H_{1/3}-1)r_0}{24 r}+\frac{\left(2r^3+r_0^3\right)^2}{24rr_0^5} \, _2F_1\left(\frac{1}{3},1,\frac{4}{3};\frac{r_0^3}{r^3}\right)
   \log \left(1-\frac{r_0^3}{r^3}\right)-\frac{4 r^6+3r_0^3 r^3-r_0^6}{72 r^2 r_0^4},\\
& f_1 = \frac{(H_{1/3}-1)r_0}{24 r}
+\frac{r_0^2}{8 r^2} \,   _2F_1\left(\frac{1}{3},1,\frac{4}{3};\frac{r_0^3}{r^3}\right)+\frac{r (2 r^3-3r_0^3)}{72r_0^4}+\frac{r_0}{24 r}
   \log \left(1-\frac{r_0^3}{r^{3}}\right)-\frac{(6 r-r_0)r_0}{72 r^2}.
\end{align}
\end{subequations}
where $_2F_1(a,b,c;x)$ is the hypergeometric function.

\section{Transition region}\label{sec:transition}
Since the GB-dominant condition (\ref{eq:GB-dominant})
breaks down near the flat region, the previous GB-dominant solution in the large $\alpha$ approximation needs a continuation to the asymptotically flat region 
through the transition region around $r\sim r_{\rm tr}=r_0\alpha^\frac{2}{n+2}$, in which 
we will use the rescaled coordinate
\begin{align}
 u := r/r_{\rm tr}.\label{eq:rescaled-u}
\end{align}
The solution in the transition region is matched with the GB-dominant solution in the inner overlap region ($r_0 \ll r \ll r_{\rm tr}$ or $u \ll 1$) and with the asymptotic solution~(\ref{eq:GR-asym}) in the outer overlap region ($r \gg r_{\rm tr}$ or $u \gg 1$), respectively.
As in the GB region, the $n>2$ cases and $n=2$ case are studied separately.

 \subsection{$n>2$}
First, we consider the behavior of the GB-dominant
solution in the inner overlap region in terms of the rescaled coordinate~(\ref{eq:rescaled-u}), in which the leading order solution~(\ref{eq:largeA-corr-sol-a1}) behaves as
\begin{align}
b/b_H \simeq  f \simeq 1 - \left(\frac{r_0}{r}\right)^\frac{n-2}{2} = 1 - \alpha^{-\frac{n-2}{n+2}} u^{-\frac{n-2}{2}},\label{eq:match-LO}
\end{align}
From eq.~(\ref{eq:app-Fx-as}), the next-to-leading order correction~(\ref{eq:largeA-corr-sol-a1}),(\ref{eq:largeA-corr-sol-b1}),(\ref{eq:largeA-corr-sol-f1}) behaves as the same order, 
\begin{align}
 \frac{a_1}{\alpha} \sim \frac{b_1}{\alpha} \sim \frac{f_1}{\alpha} 
 \sim \frac{r^2}{\alpha} = \alpha^{-\frac{n-2}{n+2}} u^2.
\end{align}
Therefore, we expect the metric functions  in the transition region to take the following form
\begin{align}
 a = a_H \left(1 + \alpha^{-\frac{n-2}{n+2}} A(u) \right),\quad  b = b_H \left(1 + \alpha^{-\frac{n-2}{n+2}} B(u) \right),\quad 
  f = 1 + \alpha^{-\frac{n-2}{n+2}} F(u),\label{eq:metric-ngtr2-trans-ansatz}
\end{align}
where $A(u),B(u),F(u)$ are regular functions of $u$, which satisfy
\begin{align}
 A(u) \simeq {\rm const}.,\quad B(u) \simeq -u^{-\frac{n-2}{2}},\quad F(u) \simeq -u^{-\frac{n-2}{2}}\quad {\rm for} \quad u \ll1.\label{eq:metric-ngtr2-match}
\end{align}
From the behavior of the metric~(\ref{eq:metric-ngtr2-trans-ansatz}), one can expect a certain simplification by expanding the EGB equation~(\ref{eq:EGBeqnABCF}) in the power of $\alpha^{-\frac{n-2}{n+2}}$.
Note that the same power $\alpha^{-\frac{n-2}{n+2}}$ also appears in the effective potential of the EGB black holes~\cite{Suzuki:2022snr}.
We can notice that eqs.~(\ref{eq:eqnA}) and (\ref{eq:eqnB}) to the leading order in $\alpha^{-\frac{n-2}{n+2}}$ are integrable to give
\begin{align}
A'(u)=\frac{(n^2-1) u^{-n-1} \left( \alpha_1 u^2+u^{n+2}F(u)   -u^n F(u)^2 \right)}{2 (n+1) F(u)-(n-1)u^2}\label{eq:dAdu},
%A'(u)=\frac{(n+1) u^{-n-1} \left( (n-1) n \,\alpha_1 u^2+u^{n+2}F(u)   -(n-1) n u^n F(u)^2 \right)}{2 n (n+1) F(u)-u^2}\label{eq:dAdu},
\end{align}
\begin{align}
B'(u)=\frac{(n^2-1) u^{-n-1} \left( \beta_1 u^2+u^{n+2}F(u)  - u^n F(u)^2 \right)}{2  (n+1) F(u)-(n-1)u^2}\label{eq:dBdu},
%B'(u)=\frac{(n+1) u^{-n-1} \left( (n-1) n \,\beta_1 u^2+u^{n+2}F(u)   -(n-1) n u^n F(u)^2 \right)}{2 n (n+1) F(u)-u^2}\label{eq:dBdu},
\end{align}
where $\alpha_1,\beta_1$ is integration constants.
Using the above equations, and eliminating $A'(u)$ and $B'(u)$ from the leading order of Eq.~(\ref{eq:eqnC}), we obtain the quartic equation with respect to $F(u)$
\begin{align}
& \left(n^2-1\right)   u^8 \left( 2 \alpha_1 \beta_1  \left(n+1\right)u^{-2n-4}+(n-1)(\alpha_1+\beta_1) u^{-n-2}\right) \nonum
&   \quad- (n-1)^2  u^{6} \left(2 (\alpha_1+\beta_1)  (n+1) u^{-n-2}-n-2 \right)F(u)\nonum
&   \quad + (n-1)u^{4} \left(2 (\alpha_1+\beta_1) (n+1)^2 u^{-n-2}-3(n-1) (n+2) \right) F(u)^2 \nonum
& \quad +4  (n^2-3)(n+1)  u^2 F(u)^3
   - 2  (n+1)^3 F(u)^4  = 0. \label{eq:4th}
\end{align}
If we introduce the function $\tilde F(u)$ and the variable $x$ by
\begin{align}
 \tilde{F} := F/u^2 ,\quad x := u^{-n-2},\label{eq:def-tF-x}
\end{align}
Eq.~(\ref{eq:4th}) reduces to the quadratic equation with respect to $x$ as
\begin{align}
 \tilde{F}(x) \left(-2  (n+1)^3 \tilde{F}(x)^3+4 (n+1)(n^2-3) \tilde{F}(x)^2-3 (n-1)^2(n+2)
   \tilde{F}(x)+(n-1)^2(n+2)\right)\nonum
   +\left(n^2-1\right)   \left(\alpha _1+\beta
   _1\right) \left(2  \left(n+1\right) \tilde{F}(x)^2-2 (n-1) 
   \tilde{F}(x)+n-1\right)  x
    +2 \alpha _1 \beta _1 \left(n-1\right)(n+1)^2  x^2=0,
\end{align}
which admits two branches
\begin{align}
 &x =  \fr{4(n-1)(n+1)^2(n^2-1) \alpha_1 \beta_1}\left[ - (\alpha_1+\beta_1)(2  (n+1) n^2\tilde{F}^2-2 (n-1)\tilde{F}  -n+1) \pm \sqrt{Q(\tilde{F})}\right],\label{eq:x-tF}
\end{align}
where
\begin{align}
& Q(\tilde{F})  =4 (n-1) (n+1)^4  \left(2(3   n+1) \alpha _1 \beta _1+(n-1)(\alpha _1^2+\beta _1^2)\right)\tilde{F}^4\nonum
&  \quad - 8 (n+1)^3 (n-1) \left(2 (n+1) (3 n-5) \alpha _1 \beta _1+ (n-1)^2(\alpha
   _1^2+\beta_1^2)\right) \tilde{F}^3
   \nonum
 & \quad +8 (n+1)^2(n-1)^3
    \left((5n+6) \alpha _1 \beta _1+n \left(\alpha _1^2+\beta _1^2\right)\right)\tilde{F}^2\nonum
& \quad  +4 (n+1)^2(n-1)^3 
   \left(\left(\alpha _1-\beta _1\right){}^2-n \left(4 \alpha _1 \beta
   _1+\alpha _1^2+\beta _1^2\right)\right)\tilde{F}
   +(n-1)^4(n+1)^2\left(\alpha _1+\beta   _1\right){}^2.
\end{align}

\subsubsection{Matching in the inner overlap region $r_0 \ll r\ll r_{\rm tr}$ ($u\ll1$)}

Now, we consider the matching between the GB region, which has an overlap with the transition region for $r_0 \ll r\ll r_{\rm tr}$ ($u\ll1$). 
The matching condition~(\ref{eq:metric-ngtr2-match}) requires
\begin{align}
 \tilde{F} \simeq -u^{-\frac{n+2}{2}} = -\sqrt{x} \quad {\rm for} \quad x \ll 1. \label{eq:tFx-ex-gb},
\end{align}
One the other hand, expanding eq.~(\ref{eq:x-tF}) for $|\tilde{F}| \gg 1$, 
 we obtain
\begin{align}
 x \simeq \frac{-(n-1)(\alpha_1+\beta_1)\pm \sqrt{(n-1)((n-1)(\alpha_1^2+\beta_1^2)+2(3n+1)\alpha_1\beta_1)}}{2(n-1)\alpha_1\beta_1} \tilde{F}^2 + \ord{\tilde{F}},
\end{align}
yielding a matching condition
\begin{align}
 1  =  -\frac{2(n-1)\alpha_1\beta_1}{(n-1)(\alpha_1+\beta_1)\pm \sqrt{(n-1)((n-1)(\alpha_1^2+\beta_1^2)+2(3n+1)\alpha_1\beta_1)}}.\label{eq:cond-f1}
\end{align}
Plugging the behavior of $F(u)$ in eq.~(\ref{eq:metric-ngtr2-match}) into eqs.~(\ref{eq:dAdu}) and (\ref{eq:dBdu}), we obtain $A$ and $B$ for $u \ll 1$,
\begin{align}
&A \simeq {\rm const.} + \frac{(n-1)(\alpha_1-1)}{n-2} u^{-\frac{n-2}{2}},\quad
B \simeq {\rm const.} + \frac{(n-1)(\beta_1-1)}{n-2} u^{-\frac{n-2}{2}}.
\end{align}
Comparing this with $A(u)$ and $B(u)$ in eq.~(\ref{eq:metric-ngtr2-match}), we can determine
\begin{align}
 \alpha_1= 1,\quad \beta_1 =\fr{n-1}.
\end{align}
It is easy to check that this condition also satisfies eq.~(\ref{eq:cond-f1}).

\subsubsection{Matching in the outer overlap region $r\gg r_{\rm tr}$ ($u\gg 1$)}

In the outer overlap region $r\gg r_{\rm tr}$ ($u\gg 1$),  matching with the asymptotically flat background $f\to 1$ requires $\tilde{F} \simeq  0$ for $x \ll 1$. Expanding eq.~(\ref{eq:x-tF}) around $\tilde{F} \simeq 0$, we obtain
\begin{align}
 x \simeq \frac{n-1}{4(n+1)\alpha_1\beta_1} \left( -\alpha_1-\beta_1 \pm |\alpha_1+\beta_1|\right)+\ord{\tilde{F}}.
\end{align}
Therefore, for the consistent match, we should choose $(+)$ branch because $\alpha_1+\beta_1=n/(n-1)>0$, that gives
\begin{align}
 x =  \fr{4n^2(n^2-1)}\left[ -n (2n^2(n^2-1) \tilde{F}^2 -2 n(n-1)\tilde{F} +1) + \sqrt{\tilde{Q}(\tilde{F})}\right]\label{eq:x-tF2}
\end{align}
with
\begin{align}
& \tilde{Q}(\tilde{F}) = 4 n^4 (n^2-1)^2(n+2)^2 \tilde{F}^4- 8 n^3 (n^2-1)(n^3+3n^2-12)\tilde{F}^3\nonum
 &\quad +8n^2(n-1)(n^3+3n^2+3n-6)\tilde{F}^2 -4 n(n-1)(n^2+2n+4)\tilde{F}+n^2.
\end{align}
Finally, from eq.~(\ref{eq:x-tF2}), we can determine the behavior of $F$  in the outer overlap region $r\gg r_{\rm tr}$  ($u\gg 1$) as
\begin{align}
 F \simeq -\frac{n(n+1)}{(n-1)(n+2)} \fr{u^n},
\end{align}
which also determines the behavior of $A$ and $B$ through eqs.~(\ref{eq:dAdu}) and (\ref{eq:dBdu}),
\begin{align}
 A \simeq  {\rm const.} -\frac{2(n+1)}{n(n-1)(n+2)} \fr{u^{n}},
 \quad  B \simeq  {\rm const.}-\frac{(n+1)(n^2-2)}{n(n-1)(n+2)} \fr{u^{n}}
\end{align}
From the original form~(\ref{eq:metric-ngtr2-trans-ansatz}),
the metric functions behave as
\begin{subequations}\label{eq:asym-ab}
\begin{align}
&  a \simeq a_H \left(1+  {\rm const.}\times \alpha^{-\frac{n-2}{n+2}} \right) - \frac{2(n+1)\alpha \, a_H }{n(n-1)(n+2)} \frac{r_0^n}{r^{n}},
% = 1 - \frac{2(n+1)\alpha r_0^{n}}{n(n-1)(n+2)} \fr{r^{n}}\left(1+\ord{\alpha^{-\frac{n-2}{n+2}}}\right)
\\
 &  b \simeq b_H \left(1 +  {\rm const.}\times\alpha^{-\frac{n-2}{n+2}}\right) -  \frac{(n+1)(n^2-2)\alpha \, b_H}{n(n-1)(n+2)} \frac{r_0^n}{r^{n}}.
 % = 1 - \frac{(n+1)(n^2-2)\alpha r_0^{n}}{n(n-1)(n+2)} \fr{r^{n}}\left(1+\ord{\alpha^{-\frac{n-2}{n+2}}}\right),
\end{align}
\end{subequations}
 The scaling constants need $\ord{\alpha^{-\frac{n-2}{n+2}}}$ corrections so that the metric asymptotes to the Minkowski at $r\to\infty$,
\begin{align}
 a_H = 1 + \ord{\alpha^{-\frac{n-2}{n+2}}},\quad b_H = 1 + \ord{\alpha^{-\frac{n-2}{n+2}}}.\label{eq:scalings}
\end{align}
From the ADM formula~(\ref{eq:adm-mass-tau}), eq.~(\ref{eq:asym-ab}) leads to the mass and tension at large $\alpha$
\begin{align}
 M \simeq \frac{(n+1)\Omega_{n+1}}{16\pi G}  \alpha\, r_0^{n} L,\label{eq:BS-mass-LO}
\end{align}
\begin{align}
{\cal T} \simeq \frac{(n+1)\Omega_{n+1}}{16\pi G(n-1)} \alpha \,r_0^{n}.
\end{align}
Particularly, the relative tension approaches a finite limit at large $\alpha$
\begin{align}
N = \frac{L {\cal T}}{\cal M} \simeq \fr{n-1},\label{eq:BS-N-LO}
\end{align}
which is greater than that of the black string in GR
\begin{align}
 N_{\rm GR} = \fr{n+1}.
\end{align}
Since the scaling constants are determined as $a_H\simeq b_H \simeq 1$,
the temperature and entropy are identical to that of the pure Gauss-Bonnet metric~(\ref{eq:pureGBgtr2}) at the leading order
\begin{align}
T_{\rm H} \simeq \frac{n-2}{8\pi} \fr{r_0},\label{eq:BS-T-LO}
\end{align}
\begin{align}
 S \simeq \frac{(n+1) \Omega_{n+1}}{2G(n-1)} \alpha\, r_0^{n+1} L \label{eq:BS-ent-LO}.
\end{align}
It is easy to verify that these variables satisfy the first law with the variation of $(r_0,L)$ at $\ord{\alpha}$
\begin{align}
 dM = T_{\rm H} dS + {\cal T} dL.
\end{align}
and the Smarr-type formula~(\ref{eq:smarr-formula})~\cite{Brihaye:2010me}
\begin{align}
 M = {\cal T} L + T_{\rm H} S.
\end{align}
Note that the matching result~(\ref{eq:scalings}) shows that the corrections to these variables should be
given in the power of $\alpha^{-\frac{n-2}{n+2}}$, or both of $\alpha^{-\frac{n-2}{n+2}}$ and $\alpha^{-1}$, rather than the simple expansion in $\alpha^{-1}$.
However, we will not pursue determining the correction terms, as it will require more laborious calculations.

\subsubsection{Fragmentation at large $\alpha$}
As we obtained the mass and entropy of the black string of the length $L$ at large $\alpha$, 
let us compare the entropy with that of the black hole of the same mass.
The mass and entropy of the $d=n+4$ Boulware-Deser black hole with the radius $r_H$ is given by
\begin{align}
 M_{\rm BH} = \frac{(n+2)\Omega_{n+2}}{16\pi G} \left(n(n+1)\alpha_{\rm GB}+r_H^2\right) r_H^{n-1} \simeq 
  \frac{(n+1)(n+2)\Omega_{n+2}}{16\pi G(n-1)} \alpha\, r_0^2 \, r_H^{n-1}
\end{align}
and
\begin{align}
 S_{\rm BH} = \frac{\Omega_{n+2}}{4 G} \left(1+\frac{2 (n+1)(n+2)\alpha_{\rm GB}}{r_H^2}\right)r_H^{n+2} \simeq
  \frac{(n+1)(n+2)\Omega_{n+2}}{2G n (n-1)} \alpha\, r_0^2\, r_H^{n}.
\end{align}
where eq.~(\ref{eq:def-alpha}) is used.
Comparing with eq.~(\ref{eq:BS-mass-LO}), the corresponding black hole radius is given by
\begin{align}
r_H =  \left(\frac{(n-1)\Omega_{n+1}}{(n+2)\Omega_{n+2}}\right)^\fr{n-1}  r_0^\frac{n-2}{n-1} L^\fr{n-1}.
\end{align}
Therefore, using eq.~(\ref{eq:BS-ent-LO}), the entropy ratio becomes
\begin{align}
 \frac{S}{S_{\rm BH}} \simeq \frac{n\Omega_{n+1}}{(n+2)\Omega_{n+2}} \frac{r_0^{n-1} L }{r_H^n}
 = C \left(\frac{r_0}{L}\right)^\fr{n-1},\quad C := \frac{n}{n-1}\left(\frac{(n+2)\Omega_{n+2}}{(n-1)\Omega_{n+1}}\right)^{\fr{n-1}}
\end{align}
This ensures that the black hole phase has larger entropy, and hence the Gregory-Laflamme instability occurs for thin enough strings $r_0 \ll L$ even at the large $\alpha$ limit,
which is in accordance with the fact that 
pure GB black strings are dynamically unstable~\cite{Giacomini:2015dwa}.

\subsection{$n=2$}

From the GB-dominant solution in eqs.~(\ref{eq:GBsol-n2-0}) and (\ref{eq:GBsol-n2-1}), the behavior in the inner overlap region $r_0\ll r \ll r_{\rm tr}$ ($u\ll1$) is given by
\begin{align}
a \simeq a_H \alpha \left(u^2-\frac{u^4}{6}\right),\quad b \simeq b_H \alpha \left(u^2-\frac{u^4}{18}\right),\quad f \simeq \frac{u^2}{6} +\frac{u^4}{36}.
\end{align}
This suggests the metric functions should take the following form in the transition region
\begin{align}
 a = a_H \alpha \, \bar{A}(u),\quad b = b_H \alpha\, \bar{B}(u),\quad f = \bar{F}(u),\label{eq:metric-n2-trans-ansatz}
\end{align}
where $\bar{A}(u), \bar{B}(u)$ and $\bar{F}(u)$ are regular functions of $u$.
Unfortunately, unlike the $n>2$ cases, this ansatz never simplifies the EGB equation~(\ref{eq:EGBeqnABCF}) in the transition region, 
since all the terms become comparable as functions of $u$.
However, we can still estimate the behavior  in the outer overlap region as
\begin{align}
 \bar{A}(u) ,  \bar{B}(u) ,\bar{F}(u) \simeq {\rm const.} + \frac{{\rm const.}}{u^2},\quad {\rm for} \quad u \gg 1.
\end{align}
Therefore, the asymptotic behavior of metric functions become
\begin{align}
& a = a_H \alpha \left({\rm const.}+ {\rm const.}u^{-2}\right) = 1 + {\rm const.} \frac{\alpha }{r^2}\\
&   b = b_H \alpha \left({\rm const.}+ {\rm const.} u^{-2}\right) = 1 + {\rm const.}\frac{\alpha }{r^2},
\end{align}
where the scaling constants are set as $a_H,b_H \sim \alpha^{-1}$ so that the Minkowski background is recovered at $r\to\infty$.
This estimates the large $\alpha$ behavior of each variables
\begin{align}
  M \sim \alpha, \quad {\cal T} \sim \alpha,\quad T_{\rm H} \sim \sqrt{b_H \alpha^{-1}} \sim \fr{\alpha},\quad S \sim \alpha \sqrt{a_H} \sim \sqrt{\alpha}.\label{eq:largeA-n2-vars}
\end{align}
Using the Smarr formula $N = 1 - T_{\rm H}S/M$, one can also estimate the behavior of the relative tension as
\begin{align}
  N = 1- \ord{ \alpha^{-3/2}}.\label{eq:largeA-n2-N}
\end{align}

\section{Comparison with Numerics}\label{sec:numeric}
To compare with the analytic formula at large $\alpha$, 
we solved eq.~(\ref{eq:EGBeqnABCF}) with the Newton-Raphson method for $n=2,\dots,6$ using the following two grids
\begin{align}
 X := \frac{r_0^n}{r^n} \label{eq:grid-1}
\end{align}
or
\begin{align}
\widetilde{X} := \frac{2 (\delta-1)^2(r/r_0)^n}{\delta(1+\delta) (r/r_0)^{2n}+(\delta^2-8\delta+3)(r/r_0)^n+3\delta-1},
 \quad \delta := \left(2+0.2 \left(\frac{n\alpha}{n-1}\right)^\frac{2n}{n+2}\right)^{-1}.\label{eq:grid-2}
\end{align}
The former is used for the $n=2$ case and for small value of $\alpha$ in other dimensions ($n=3,4,5,6$).
For $n>2$, the latter is used to keep the resolution around the transition region at large $\alpha$. We used $100,200,400$ or $800$ meshes with the fourth order difference scheme, so that the Smarr formula~(\ref{eq:smarr-formula}) satisfies enough accuracy.

\begin{figure}[H]
\centering{
\includegraphics[width=7.4cm]{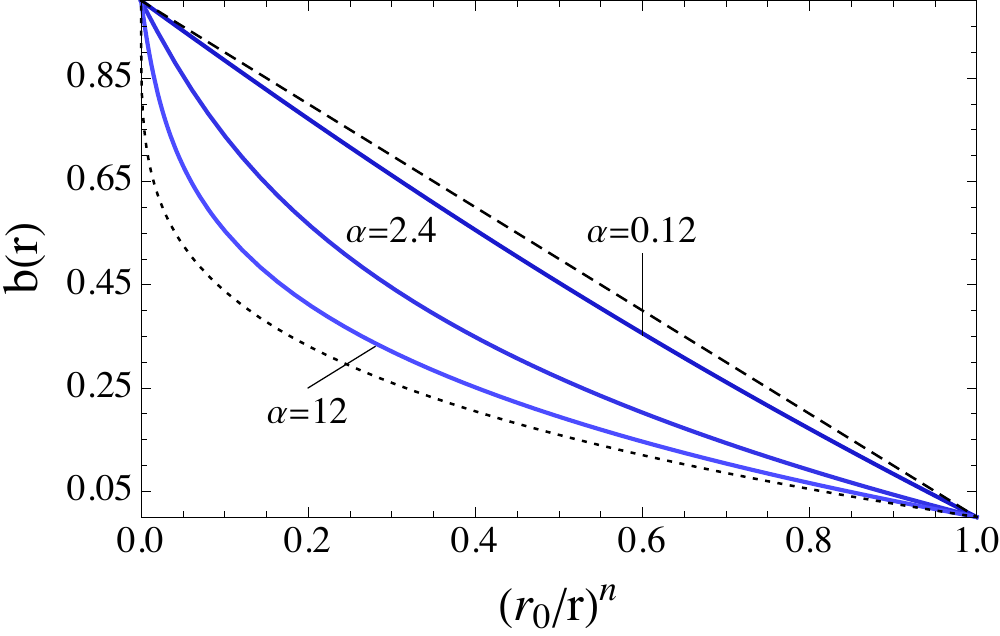}\hspace{0.3cm}
\includegraphics[width=7.4cm]{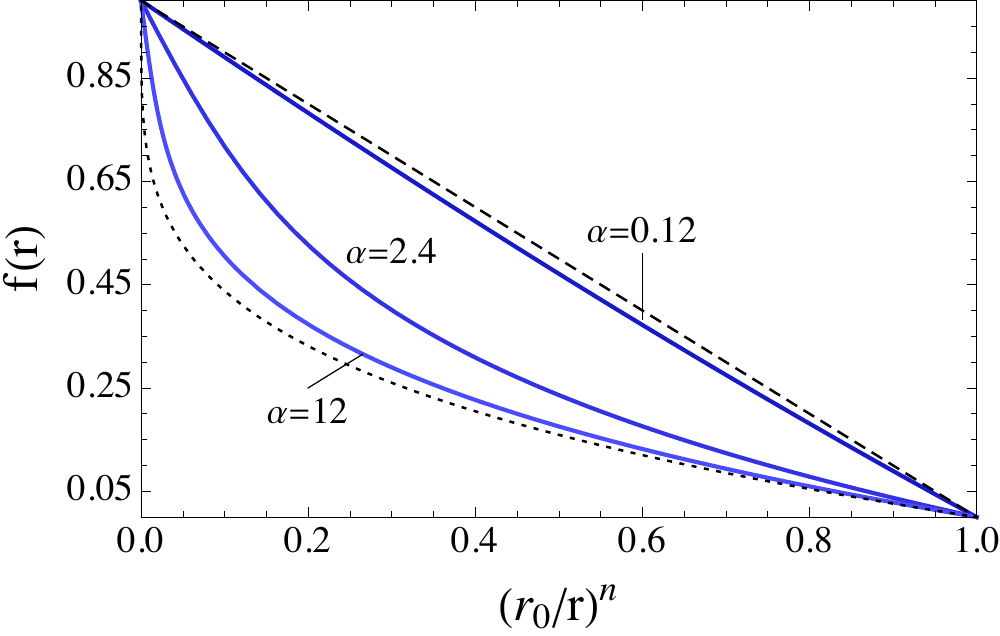}
\includegraphics[width=7.4cm]{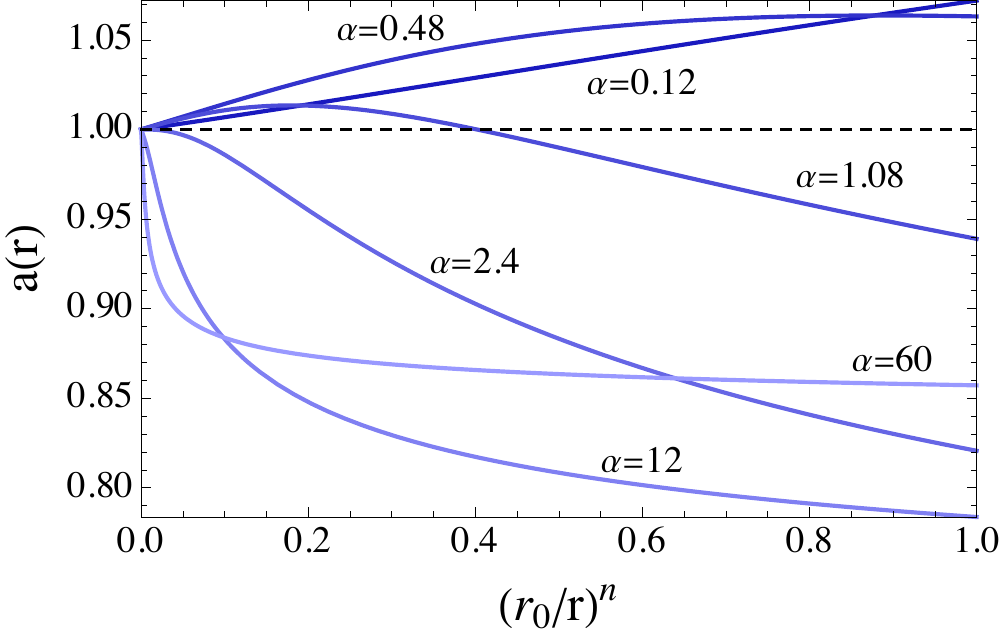}\hspace{0.3cm}
\includegraphics[width=7cm]{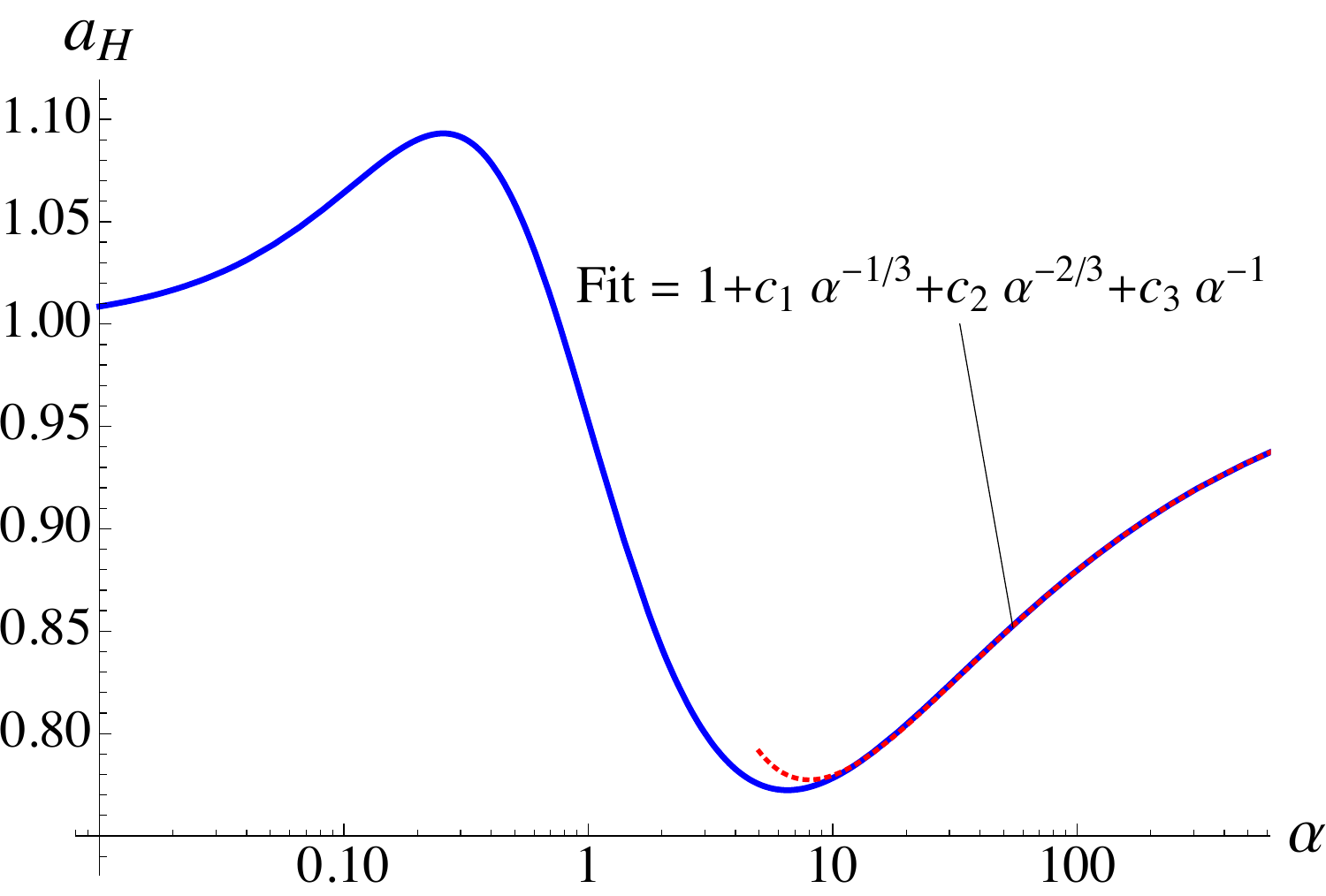}
\caption{Metric functions in $n=4$ ($d=8$). GR and pure GB black strings correspond to the black dashed and dotted curves respectively. $a_H=a(r_0)$ admits an oscillatory behavior as a function of $\alpha$, which eventually goes back to $a_H=1$ in $\ord{\alpha^{-\frac{n-2}{n+2}}}$. The red dotted curve is a fit curve. \label{fig:metricplots}}
}\end{figure}

\begin{figure}[H]
\centering{
\includegraphics[width=7.3cm]{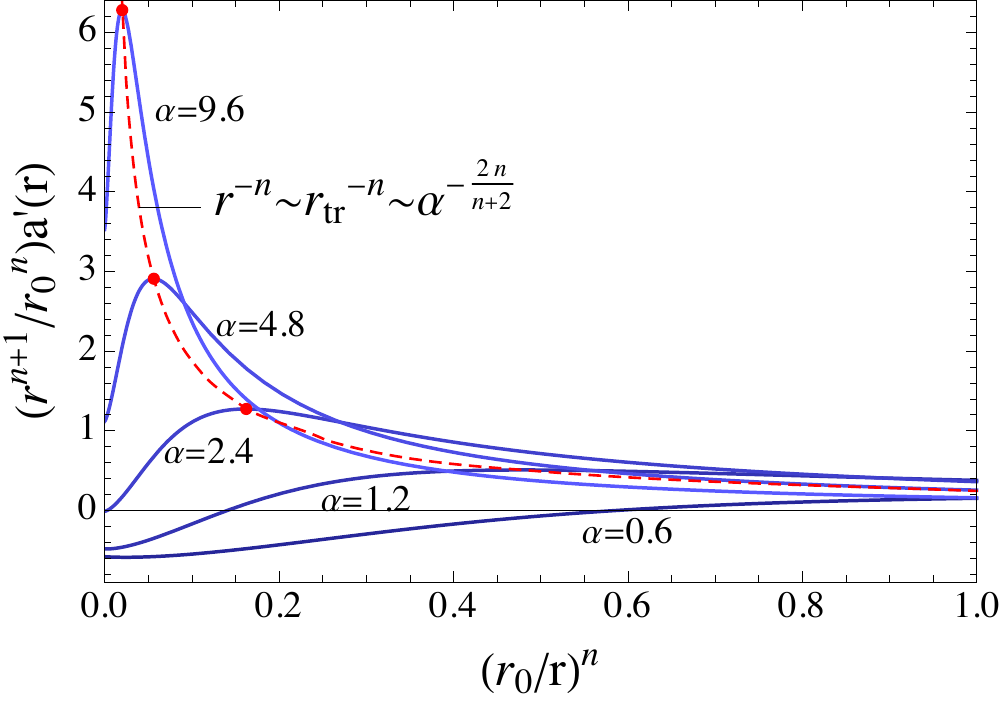}\hspace{0.5cm}
\includegraphics[width=7.3cm]{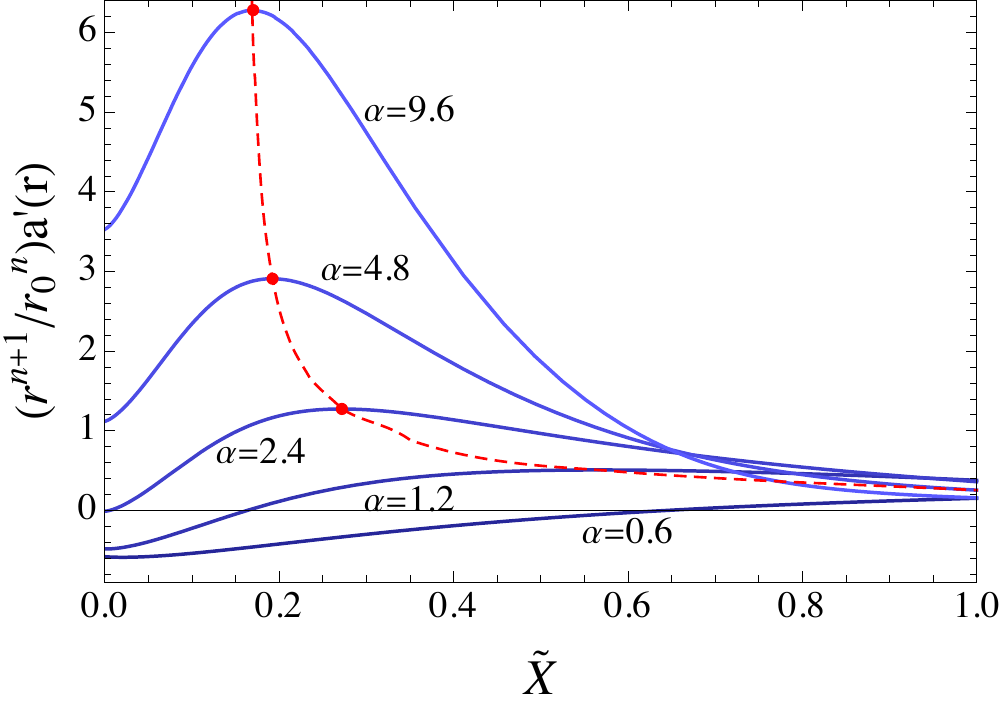}
\caption{Appearance of the transition scale $r_{\rm tr}$ in the metric function. The red dashed curves are the peak position for each $\alpha$. In the right panel, one can see that another coordinate $\widetilde{X}$ keeps the transition region in the center area for better resolution at larger $\alpha$.\label{fig:daplot}}}
\end{figure}

In the fig.~\ref{fig:metricplots}, the metric functions for $n=4$ are presented. The metric approaches to that of the pure GB black strings at large $\alpha$. The appearance of the transition region is clearer if we take a look on the quantity $r^{n+1}r_0^{-n} a'(r) \propto a'(X)$ (fig.~\ref{fig:daplot}). 

\begin{figure}[H]
\centering{
\includegraphics[width=7.5cm]{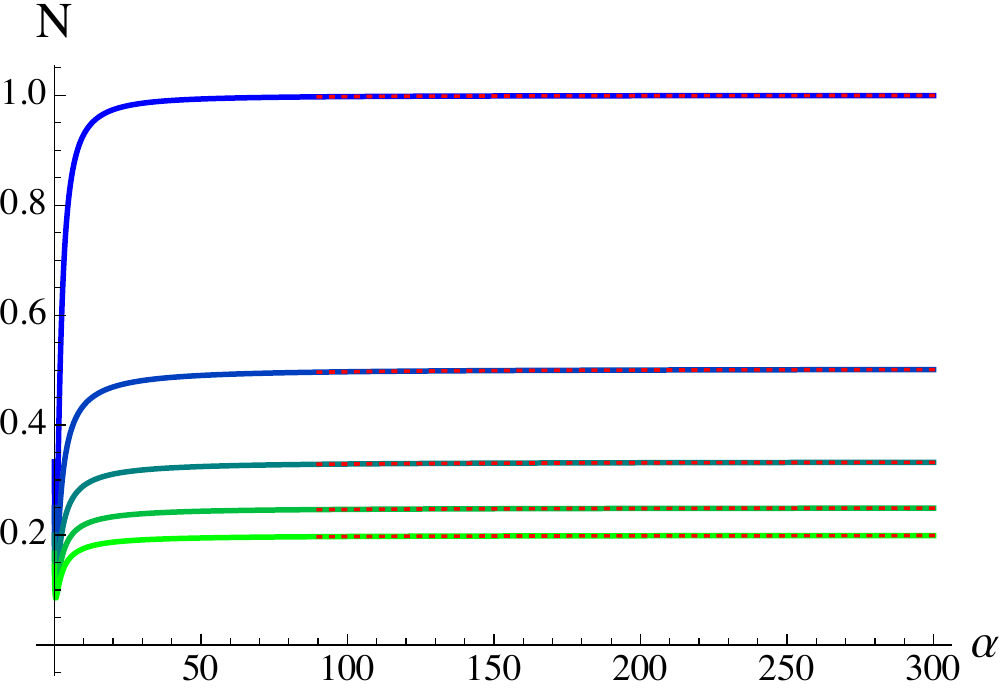}
\includegraphics[width=8.5cm]{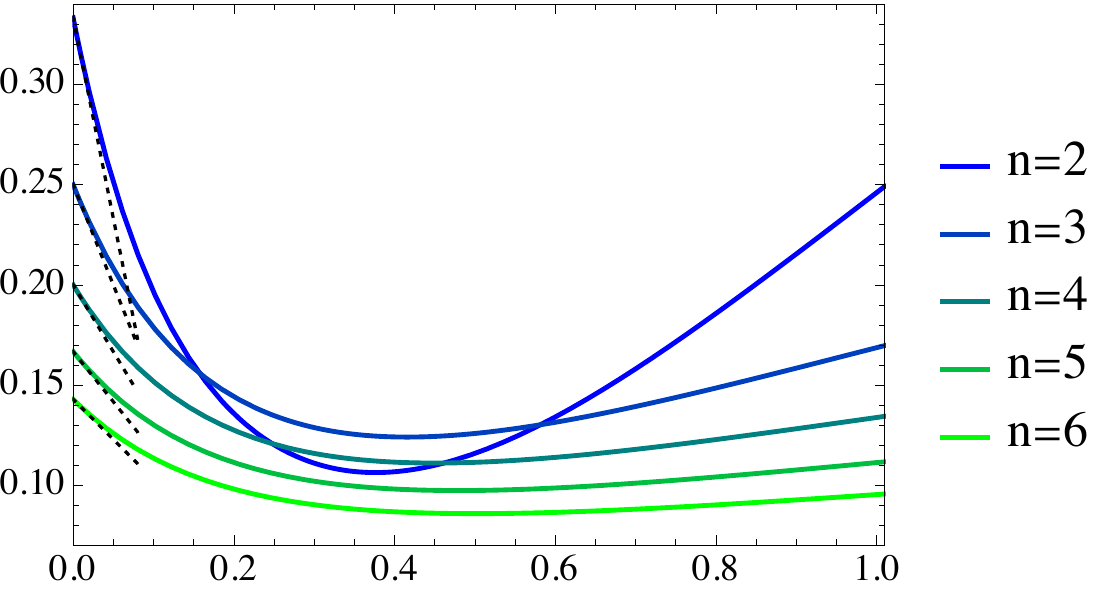}
\caption{Relative tension v.s. $\alpha$. Numerical results are plotted by solid curves for each dimension. Red dotted curves are the fitting curves in eq.~(\ref{eq:fit-N}) for $n>2$ and eq.~(\ref{eq:fit-n2}) for $n=2$. The right panel is a closeup of the small parameter region. Black dotted curves are the linear approximation~(\ref{eq:N-lin}). \label{fig:Nplot}}
}
\end{figure}

\begin{figure}[H]
\centering{
\includegraphics[width=7.5cm]{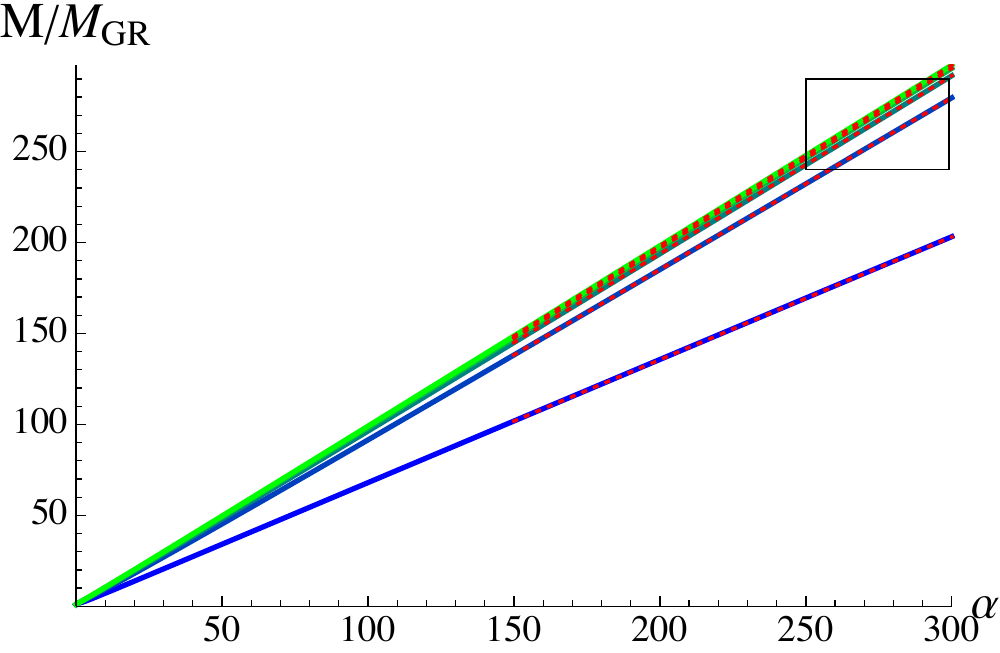}
\includegraphics[width=8.5cm]{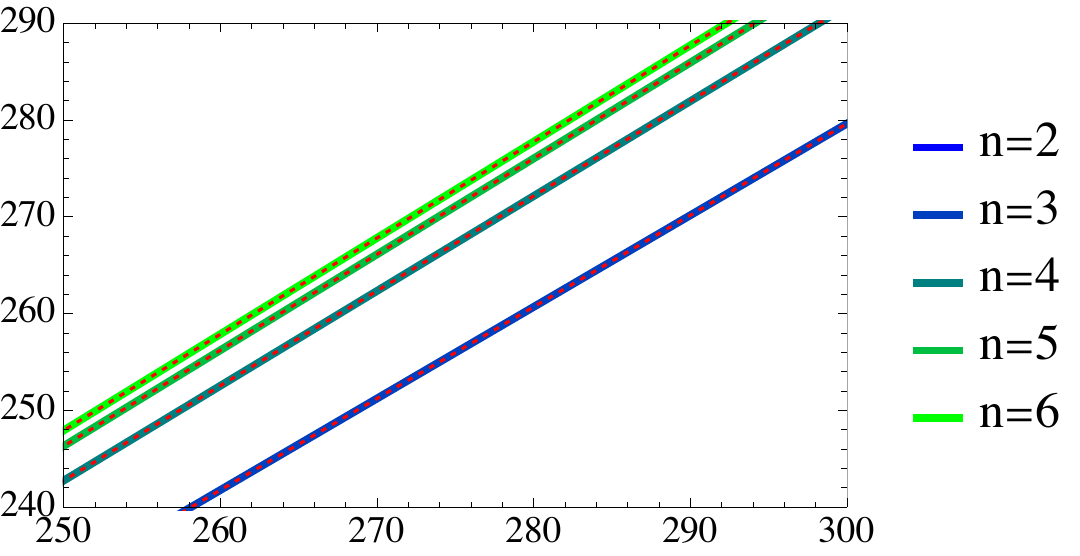}
\caption{Mass v.s. $\alpha$. Numerical results are plotted by solid curves for each dimension. Red dotted curves are the fitting curves in eq.~(\ref{eq:fit-M}) for $n>2$ and  eq.~(\ref{eq:fit-n2}) for $n=2$.\label{fig:Mplot}}
}\end{figure}

\begin{figure}[H]
\centering{
\includegraphics[width=7.5cm]{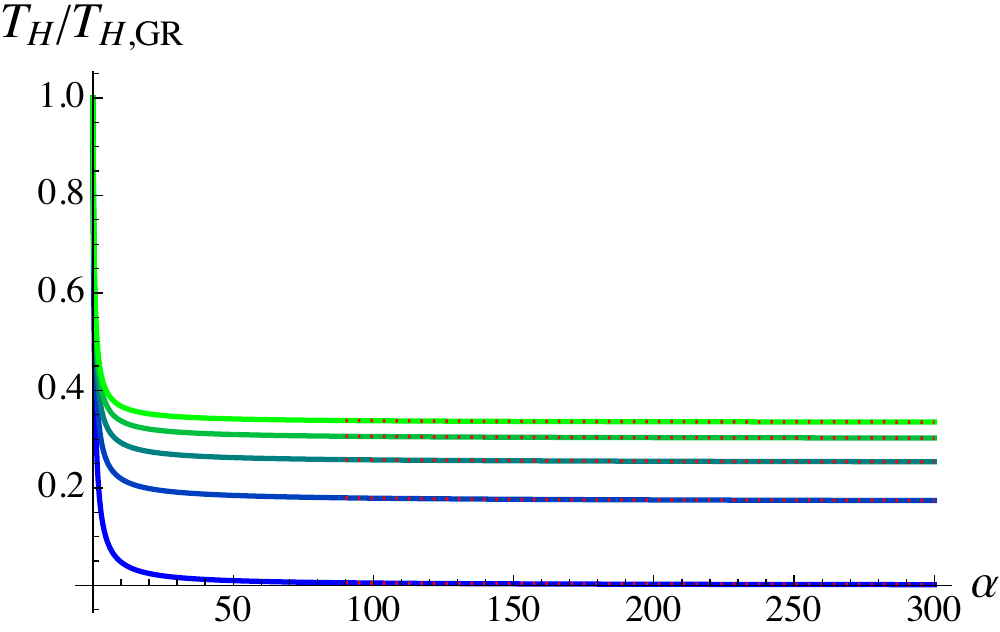}
\includegraphics[width=8.8cm]{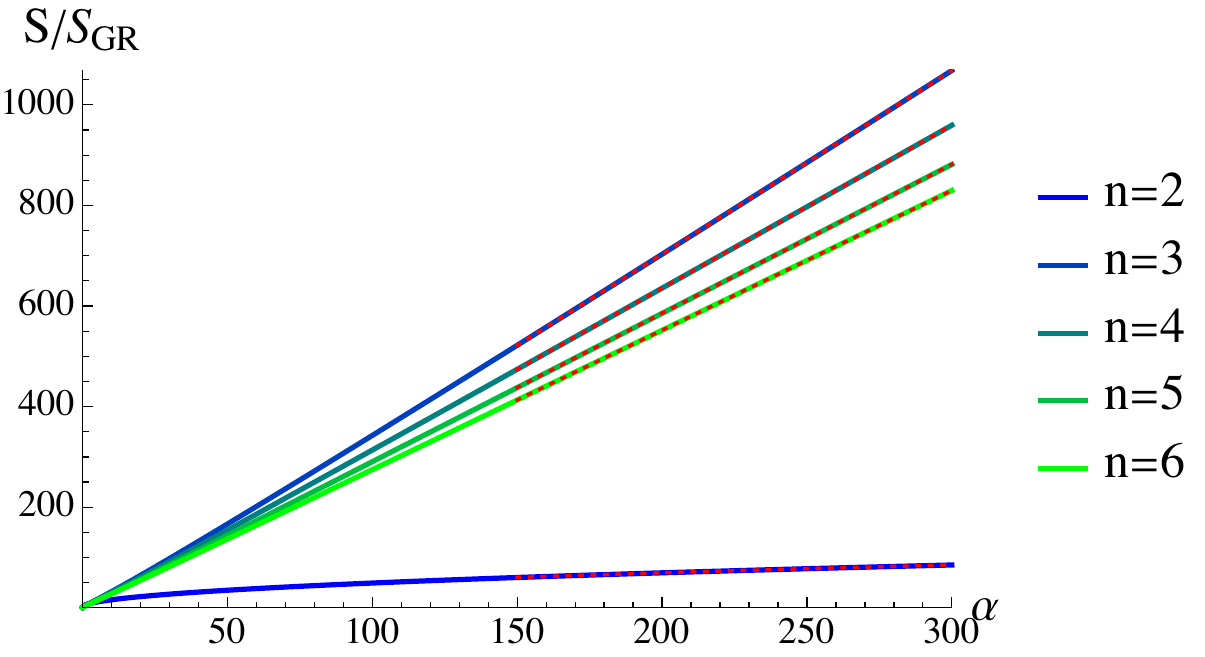}
\caption{Temperature and entropy v.s. $\alpha$. Numerical results are plotted by solid curves for each dimension.
Red dotted curves are the fitting curves in eqs.~(\ref{eq:fit-T}) and (\ref{eq:fit-S}) for $n>2$ and eq.~(\ref{eq:fit-n2}) for $n=2$. \label{fig:Tplot}}
}
\end{figure}

In figs.~\ref{fig:Nplot}-\ref{fig:Tplot}, physical quantities of EGB black strings are compared with the large $\alpha$ results. Other than the relative tension, each quantities are normalized by the GR values~(\ref{eq:thermovars-GR}). As a characteristic behavior, one can notice that the relative tension once falls to a minimum before it gradually grows to the large $\alpha$ limit.\footnote{The same behavior has already been observed in $d=6,8$ ($n=2,4$) (See fig.~5 in ref.~\cite{Brihaye:2010me})}
Because of the slow convergence due to the fractional power $\alpha^{-\frac{n-2}{n+2}}$ in the correction, it is difficult to see the behavior at $\alpha\to\infty$ in eqs.~(\ref{eq:BS-mass-LO}), (\ref{eq:BS-N-LO}), (\ref{eq:BS-T-LO}) and (\ref{eq:BS-ent-LO}), directly. Instead, we use the following fitting curves to compare with the numerical result
\begin{align}
& N = \fr{n-1} \left(1+ \sum_{i=1} c_i \alpha^{-i\frac{n-2}{n+2}}\right),
\label{eq:fit-N}\\
& \frac{M}{M_{\rm GR}} = \alpha \left(1+ \sum_{i=1} c_i \alpha^{-i\frac{n-2}{n+2}}\right),\label{eq:fit-M}\\
& \frac{T_{\rm H}}{T_{\rm H,GR}} = \frac{n-2}{2n} \left(1+ \sum_{i=1} c_i \alpha^{-i\frac{n-2}{n+2}}\right),\label{eq:fit-T}\\
&  \frac{S}{S_{\rm GR}} = \frac{2(n+1)}{n-1}\alpha \left(1+ \sum_{i=1} c_i \alpha^{-i\frac{n-2}{n+2}}\right),\label{eq:fit-S}
 \end{align}
 where we include correction terms up to $\ord{\alpha^{-1}}$.
 The above estimates are only for $n>2$.
 For $n=2$, reflecting the results~(\ref{eq:largeA-n2-vars}) and (\ref{eq:largeA-n2-N}), we rather fit by
 \begin{align}
  N = 1 -c \, \alpha^{-3/2},\quad M = c\, \alpha,\quad S = c\, \sqrt{\alpha},\quad
  T_{\rm H} = \frac{c}{\alpha}.\label{eq:fit-n2}
 \end{align}
We find that all these fittings match the numerical results well at the large $\alpha$ region.

\section{Extension to Einstein-Lovelock black holes}\label{sec:lovelockbh}
So far, we have investigated the large $\alpha$ limit in the EGB theory.
Here, we discuss the possible extension to the Einstein-Lovelock theories~\cite{Garraffo:2008hu} whose action is given by
\footnote{For simplicity, we only consider the asymptotically flat background. 
However, we expect other non-flat backgrounds such as (A)dS or squashed Kaluza-Klein also admit the similar simplification as long as the typical scale of the background is sufficiently larger than the transition scale.
}
\begin{align}
 S = \fr{16\pi G}\int dx^d \sqrt{-g} \left (R + \sum_{k=2}^{\lfloor (d-1)/2 \rfloor} \alpha_k' {\cal L}_k\right),
\end{align}
where the $k$ th Lovelock term ${\cal L}_k$ is given by
\begin{align}
{\cal L}_k := 2^{-k} \delta^{a_1b_1\cdots a_k b_k}_{c_1d_1\cdots c_k d_k}R^{c_1d_1}{}_{a_1b_1}\cdots R^{c_k d_k}{}_{a_k b_k}.
\end{align}
In $d=n+3$ dimension, the static black hole solution can be obtained by the ansatz~\cite{Wheeler:1985qd,Myers:1988ze}
\begin{align}
 ds^2 = - (1-r^2 \psi(r)) dt^2 + \frac{dr^2}{1-r^2\psi(r)}+r^2 d\Omega_{n+1}^2,
\end{align}
where $\psi(r)$ is given by the real root of the following polynomial
\begin{align}
 \frac{m_0}{r^{n+2}} = \psi + \sum_{k=2}^{\lfloor n/2+1 \rfloor} \alpha_k \psi^k,\quad \alpha_k:= \alpha_k' \prod_{\ell=3}^{2k}(n+3-\ell) .\label{eq:lovelock-bhcond}
\end{align}
The mass parameter $m_0$ is determined by
\begin{align}
  m_0 
  = r_0^n \left(1+ \sum_{k=2}^{\lfloor n/2+1\rfloor} \widehat{\alpha}_k \right),
  \quad \widehat{\alpha}_k := r_0^{-2(k-1)} \alpha_k,
\end{align}
where $\psi(r_0)=r_0^{-2}$, and $r=r_0$ is the horizon radius.
With the normalized metric function $\widehat{\psi}:= r_0^2 \psi$, the condition~(\ref{eq:lovelock-bhcond}) is rewritten
 in the dimensionless form
\begin{align}
  0 = \widehat{\psi} - \left(\frac{r_0}{r}\right)^{n+2}+ \sum_{k=2}^{\lfloor n/2+1\rfloor} \widehat{\alpha}_k \left(\widehat{\psi}{}^k - \left(\frac{r_0}{r}\right)^{n+2}\right).
\end{align}
If the $k$-th order is exclusively dominant, the solution becomes that of pure $k$-th Lovelock theory~\cite{Crisostomo:2000bb}
\begin{align}
 \widehat{\psi} \simeq \left(\frac{r_0}{r}\right)^\frac{n+2}{k} \quad \Rightarrow \quad g_{tt} \simeq -1 + \left(\frac{r_0}{r}\right)^\frac{n-2k+2}{k}.\label{eq:pureLovelockf}
\end{align}

\subsection{With a ${\cal L}_k$}
\begin{figure}
\centering{
\includegraphics[width=14cm]{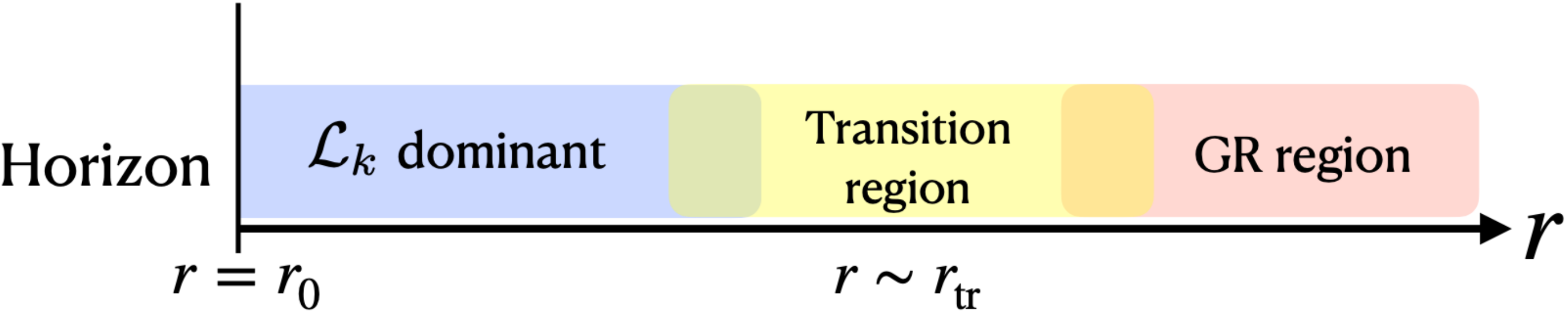}
\caption{Separate regions in the Einstein-Lovelock black hole  with a correction ${\cal L}_k$ for $\widehat{\alpha}_k\gg 1$.\label{fig:sch-Lk}}
}
\end{figure}

First, we examine whether the similar separation of scales occurs in the Einstein-Lovelock theory with a single Lovelock term ${\cal L}_k$
by assuming $\widehat{\alpha}_k \gg 1$. 
It is obvious that the Lovelock term becomes dominant around the horizon where the metric almost becomes the pure Lovelock solution~(\ref{eq:pureLovelockf}), 
%while it  admits the post-Minkowski behavior at large $r$.
while the post-Minkowski behavior is obtained at large $r$.
In the intermediate region if it exists, the following terms should be comparable in eq.~(\ref{eq:lovelock-bhcond})
\begin{align}
  \widehat{\psi}  \sim  \widehat{\alpha}_k \widehat{\psi}^k  \sim   \widehat{\alpha}_k\left(\frac{r_0}{r}\right)^{n+2},
\end{align}
which determines a transition scale
\begin{align}
r \sim r_{\rm tr} := r_0\, \widehat{\alpha}_k{}^\frac{k}{(k-1)(n+2)}. 
\end{align}
Therefore, the Einstein-Lovelock theory with a single Lovelock term can admits the similar structure as in the EGB theory (see Fig.~\ref{fig:sch-Lk}).
One can check that the transition scale in the EGB theory~(\ref{eq:trans_scale}) is reproduced by setting $k=2$.

\subsection{With ${\cal L}_2$ and ${\cal L}_3$}
\begin{figure}
\centering{
\includegraphics[width=16cm]{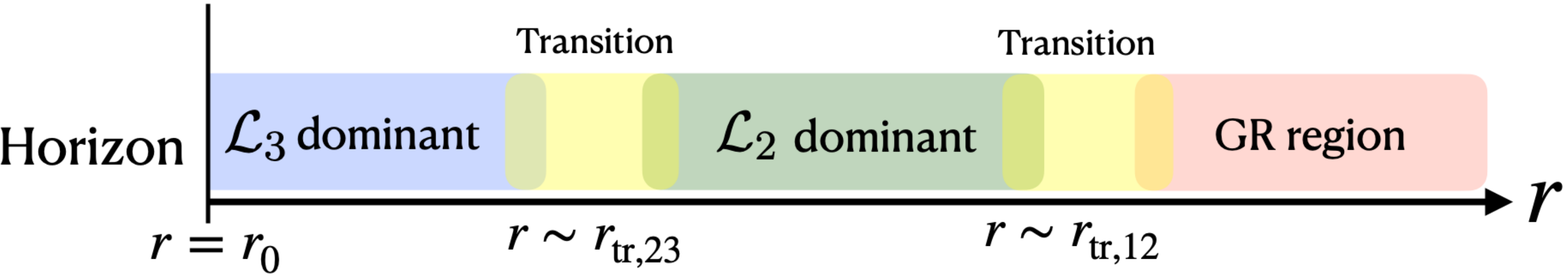}
\caption{Three separate regions in the Einstein-Lovelock black hole with ${\cal L}_2$ and ${\cal L}_3$ corrections for $1\ll \widehat{\alpha}_2 \ll \widehat{\alpha}_3 \ll \widehat{\alpha}_2^2$}.
\label{fig:sch-L2L3}
}
\end{figure}

Next, we consider more general cases in which the theory includes multiple Lovelock corrections.
For instance we focus on the Einstein-Lovelock theory only with the second order and third order Lovelock terms, whose normalized coupling constants are large:
\begin{align}
1\ll \widehat{\alpha}_2 , \widehat{\alpha}_3.
\end{align}
Here we do not assume the hierarchy between the two.
Then, the transition between the ${\cal L}_3$-dominant and ${\cal L}_2$-dominant regions can exist if the following terms become comparable in eq.~(\ref{eq:lovelock-bhcond})
\begin{align}
 \widehat{\alpha}_2 \widehat{\psi}^2 \sim  \widehat{\alpha}_3 \widehat{\psi}^3 \sim (\widehat{\alpha}_2+\widehat{\alpha}_3) \left(\frac{r_0}{r}\right)^{n+2},
\end{align}
which gives a transition scale
\begin{align}
 r \sim r_{{\rm tr},23} := r_0 \left(\widehat{\alpha}_3^2(\widehat{\alpha}_2+\widehat{\alpha}_3)/\widehat{\alpha}_2^3\right)^\frac{1}{n+2}.
\end{align}
Similarly, the transition between ${\cal L}_2$ dominant and GR region would occur if
\begin{align}
 \widehat{\psi} \sim  \widehat{\alpha}_2 \widehat{\psi}^2 \sim (\widehat{\alpha}_2+\widehat{\alpha}_3) \left(\frac{r_0}{r}\right)^{n+2}.
\end{align}
This introduces another transition scale
\begin{align}
   r \sim r_{\rm tr,12}:=r_0 \left(\widehat{\alpha}_2 (\widehat{\alpha}_2+\widehat{\alpha}_3)\right)^\frac{1}{n+2}.
\end{align}
To obtain the separation of scales $r_0 \ll r_{\rm tr,23} \ll r_{\rm tr,12}$, we also require
\begin{align}
  \frac{r_{\rm tr, 12}}{r_{\rm tr,23}} \sim \left(\frac{\widehat{\alpha}_2^2}{\widehat{\alpha}_3}\right)^\frac{2}{n+2} \gg 1,\quad 
  \frac{r_{\rm tr,23}}{r_0} = \left(\frac{\widehat{\alpha}_3^2}{\widehat{\alpha}_2^2}+\frac{\widehat{\alpha}_3^3}{\widehat{\alpha}_2^3}\right)^\frac{1}{n+2} \gg 1,
\end{align}
%which imposes another further scaling conditions
which is equivalent to
\begin{align}
  \widehat{\alpha}_2 \ll \widehat{\alpha}_3 \ll  \widehat{\alpha}_2^2.
\end{align}
This introduces an additional hierachy between $\widehat{\alpha}_2$ and $\widehat{\alpha}_3$.
With the latter condition, one can check that the discarded terms in eq.~(\ref{eq:lovelock-bhcond}) becomes negligible at each transition scale
\begin{align}
 \left.  \frac{\widehat{\psi}}{\widehat{\alpha}_2 \widehat{\psi}^2} \right|_{r=r_{\rm tr,23}} \sim \frac{\widehat{\alpha}_3}{\widehat{\alpha}_2^2} \ll 1 ,\quad
 \left.  \frac{\widehat{\alpha}_3\widehat{\psi}^3}{\widehat{\alpha}_2 \widehat{\psi}^2} \right|_{r=r_{\rm tr,12}} \sim \frac{\widehat{\alpha}_3}{\widehat{\alpha}_2^2} \ll 1.
\end{align}
Therefore, we conclude that, at the large coupling limit,
multiple Lovelock terms lead to the layered structure in which each Lovelock term becomes dominant one by one through the multiple transition regions ( Fig.~\ref{fig:sch-L2L3}).
Note that the ${\cal L}_2$-dominant region cannot appear inside the ${\cal L}_3$-dominant region, as it requires two contradicting conditions $\widehat{\alpha}_3 \ll \widehat{\alpha}_2^2$ and $\widehat{\alpha}_3 \gg \widehat{\alpha}_2^2$.

\section{Summary}\label{sec:summary}
In this article, we have constructed the analytic solutions of black strings in the $d=n+4$ EGB theory, using the novel method of the  large $\alpha$ approximation, where $\alpha$ is defined as the dimensionless GB coupling constant normalized by the horizon radius.
The points of this method are summarized as follows:
\begin{itemize}
\item For sufficiently large $\alpha$, {\it the GB region}, where the GB correction is dominant over the Einstein-Hilbert term, appears near the horizon. 
In the GB region, the black string metric can be obtained analytically by expanding in $1/\alpha$.
\item Since the spacetime is asymptotically flat in the transverse direction to the horizon, the GB correction ceases to be dominant at large enough distance from the horizon where we also have {\it the GR region} in which the metric is approximated as the post-Minkowski spacetime of GR.
\item There is the transition region between the two regions at the scale $r_{\rm tr}=r_0 \,\alpha^\frac{2}{n+2}$ at large $\alpha$.%, which first was pointed out in our previous geodesic study of the spherical EGB black hole~\cite{Suzuki:2022snr}.
\item These three regions admit overlaps at large $\alpha$, and hence the entire geometry can be obtained by the matched asymptotic expansion. 
\end{itemize}
\noindent Using this method, we have obtained the analytic solutions of black strings, from which we  have also obtained the phase diagram of the EGB black string analytically for large $\alpha$.
By solving  the EGB equations numerically using the Newton-Raphson method for $n=2,\dots,6$,
we have shown that the resulting phase diagram is consistent with the analytic formula obtained by the large $\alpha$ approach.
Lastly, we have discussed possible extensions to Einstein-Lovelock  theories.

\medskip
This work has several possibilities of development.
A quick application will be the extension to the Einstein-Lovelock black strings with a single or multiple Lovelock terms.
As seen in the last section, the analysis with a single Lovelock correction will be almost parallel to the EGB theory. 
For the theories with more than one Lovelock terms, one has to solve multiple layers around the horizon in which each Lovelock term becomes dominant in turn. 
It would be also interesting if one can apply to more realistic cases such as quantum corrected black holes in M-theory~\cite{Hyakutake:2016sig}.

\medskip
The application to a rotating black hole in the EGB theory or Einstein-Lovelock theories %a rotating EGB black hole or a \red{rotating} Einstein-Lovelock black hole
is a challenging but fruitful project, since an exact solution of such a black hole is not yet found  in the higher curvature theory (see Ref.~\cite{Suzuki:2022apk} on a rotating EGB black hole at large $d$). 
If we intend to construct an analytic solution of a rotating black hole at large $\alpha$, we first have to obtain the rotating black hole solutions in the pure GB theory or pure Lovelock theory. 
This deserves our future work.

\medskip

We should note that spacetimes in Einstein-Lovelock theories generically suffer from pathologies in the strong field regime, such as short scale instabilities~\cite{Konoplya:2008ix}, shockwave formation~\cite{Reall:2014sla}, and more remarkably the loss of hyperbolicity~\cite{Reall:2014pwa,Papallo:2017qvl,Kovacs:2020ywu}.
Nevertheless, it is not clear whether or not every solution obtained by our approach is in this pathological regime. For example, it is shown that static EGB black holes admit such instabilities and the loss of hyperbolicity at large $\alpha$ only in $d=5,6$~\cite{Konoplya:2008ix,Reall:2014pwa}, that correspond to the transverse section of $d=6,7$ black strings.
This might imply that, in higher dimensions, those pathologies become milder outside the horizon.

\medskip
Lastly, we would also like to point out the possibility of more general and sophisticated formulation.
The large $d$ limit~\cite{Emparan:2020inr} which is another successful approximation, has lead to the effective theory approach~\cite{Emparan:2015hwa} or equivalent membrane paradigm~\cite{Bhattacharyya:2015dva},
that greatly simplified the analysis. 
If one can find an effective description in the large $\alpha$ approximation, that may enhance the understandings and broaden the applicability of this approximation.

\section*{Acknowledgement}
We thank Harvey S. Reall and Norihiro Tanahashi for useful comments and discussions on the well-posedness in Lovelock theories.
This work is supported by Toyota Technological Institute Fund for Research Promotion A.
RS was supported by JSPS KAKENHI Grant Number~JP18K13541.
ST was supported by JSPS KAKENHI Grant Number 21K03560.

\appendix

\section{Equations}
Here we present the explicit form of the field equation
\begin{align}
 {\cal E}_{\mu\nu} = R_{\mu\nu} -\fr{2}R g_{\mu\nu} + \alpha_{\rm GB} H_{\mu\nu}.\nonumber
\end{align}
\paragraph{${\cal E}_{tt}$}
\begin{subequations}\label{eq:EGBeqnABCF}
\begin{align}
&0= \frac{n (n+1) a(r) (f(r)-1)}{ f(r)}+r f'(r) \left(\frac{r a'(r)}{2 f(r)}+\frac{(n+1)
   a(r)}{f(r)}\right)+(n+1)r a'(r)-\frac{r^2 a'(r)^2}{2
   a(r)}+r^2 a''(r)\nonum
&   +\frac{n(n+1)\alpha_{\rm GB}}{r^2}  \left(
   \frac{ (f(r)-1) r^2 (a'(r)^2-2a(r)a''(r))}{a(r)}
   -\frac{(n-1)(n-2) a(r)   (f(r)-1)^2}{f(r)}\right.\nonum
 &\left. \hspace{2.5cm}
 -\frac{ (3 f(r)-1)r^2 f'(r) a'(r)}{ f(r)} -\frac{2 (n-1) (f(r)-1)r (a(r) f'(r)+a'(r) f(r))}{ f(r)}
 \right).\label{eq:eqnA}
\end{align}

\paragraph{${\cal E}_{zz}$}
\begin{align}
&0= \frac{n (n+1) b(r) (f(r)-1)}{f(r)}+r f'(r) \left(\frac{r b'(r)}{2 f(r)}+\frac{(n+1)
   b(r)}{f(r)}\right)+(n+1)r  b'(r)-\frac{r^2 b'(r)^2}{2
   b(r)}+r^2 b''(r)\nonum
&   +\frac{n(n+1)\alpha_{\rm GB}}{r^2}  \left(
   \frac{ (f(r)-1)r^2 (b'(r)^2-2b''(r))}{ b(r)}
   -\frac{(n-1)(n-2) b(r)   (f(r)-1)^2}{ f(r)}\right.\nonum
 &\left. \hspace{2.5cm}
 -\frac{ (3 f(r)-1)r^2 f'(r) a'(r)}{f(r)} -\frac{2 (n-1) (f(r)-1) r(b(r) f'(r)+b'(r) f(r))}{ f(r)}
 \right).\label{eq:eqnB}
\end{align}
\paragraph{${\cal E}_{rr}$}
\begin{align}
&0=-\frac{r^2 a'(r) b'(r)}{4 a(r)   b(r)}-\frac{(n+1) r a'(r)}{2a(r)}
-\frac{(n+1) r b'(r)}{2b(r)}-\frac{n (n+1) (f(r)-1)}{2 f(r)}\nonum
   &+\frac{n(n+1)\alpha_{\rm GB}}{r^2}  \left( \frac{ (3 f(r)-1)r^2 a'(r)b'(r)}{2 a(r)
   b(r)}+(n-1)  (f(r)-1)r\left(\frac{a'(r)}{a(r)}+\frac{b'(r)}{b(r)}\right)
   +\frac{(n-2) (n-1)  (f(r)-1)^2}{2  f(r)}\right)\label{eq:eqnC}
\end{align}

\paragraph{$ {\cal E}^t{}_t+ {\cal E}^z{}_z-{\cal E}_\Omega$}
\begin{align}
&0=   \frac{r a'(r)}{a(r)}-\frac{r^2 a'(r)b'(r)}{2 a(r) b(r)}+\frac{r b'(r)}{
   b(r)}+\frac{(n+2) r f'(r)}{ f(r)}+\frac{n (n+3) (f(r)-1)}{ f(r)}\nonum
&+\frac{\alpha_{\rm GB} }{r^2} \left[  \frac{3 n r^3 f'(r) a'(r) b'(r)}{ a(r) b(r)}
+\frac{2n r^2 f'(r)  (1-3 f(r))}{  f(r)}\left(\frac{a'(r)}{a(r)}+\frac{b'(r)}{b(r)}\right)
-\frac{n   (n-1)(n-2)(n+5) (f(r)-1)^2}{ f(r)}
 \right.\nonum
&   \left.-\frac{n r^3 f(r) a'(r) b'(r)}{ a(r) b(r)} \left(\frac{a'(r)}{a(r)}+\frac{b'(r)}{b(r)} \right)
   +\frac{(n-1) n   (3 f(r)-1) r^2 a'(r) b'(r)}{ a(r) b(r)}
   + \frac{2 n f(r) r^3 (a''(r) b'(r)+a'(r)b''(r))}{a(r) b(r)}\right.\nonum
& \left.  +2n(f(r)-1)r^2\left(\frac{a'(r)^2}{a(r)^2}+\frac{b'(r)^2}{b(r)^2}-\frac{3(n-1)}{r}\left(\frac{a'(r)}{a(r)}+\frac{b'(r)}{b(r)}\right)-\frac{2a''(r)}{a(r)}-\frac{2b''(r)}{b(r)}
- \frac{(n-1)(n+4)f'(r)}{rf(r)}\right) \right]\label{eq:eqnF}
\end{align}
\end{subequations}
As shown in ref.~\cite{Brihaye:2010me}, the following combinations are integrable
\begin{align}
& R^t{}_t + \alpha \left(H^t{}_t + \fr{2} {\cal L}_{\rm GB}\right) = \fr{r^{n+1}} \sqrt{\frac{f}{ab}} \frac{d}{dr} \left[\frac{r^{n+1} b'}{2}\sqrt{\frac{af}{b}}\left(-1+\frac{2(n+1)\alpha_{\rm GB}}{r^2}\left(n(f-1)+\frac{rfa'}{a}\right)\right)\right].\\
& R^z{}_z + \alpha \left(H^z{}_z + \fr{2} {\cal L}_{\rm GB}\right) = \fr{r^{n+1}} \sqrt{\frac{f}{ab}} \frac{d}{dr} \left[\frac{r^{n+1} a'}{2}\sqrt{\frac{bf}{a}}\left(-1+\frac{2(n+1)\alpha_{\rm GB}}{r^2}\left(n(f-1)+\frac{rfb'}{b}\right)\right)\right].
\end{align}
Thus, the condition $R^t{}_t-R^z{}_z + \alpha (H^t{}_t -H^z{}_z)=0$ leads to the Smarr-type formula
\begin{align}
 M = {\cal T} L + T_{\rm H} S. \label{eq:smarr-formula}
\end{align}

\subsection{$1/\alpha$ expansion}\label{app:exp-eqn}
We define the dimensionless coupling constant $\alpha$ with the horizon radius $r_0$ as in eq.~(\ref{eq:def-alpha}).
The equation for the $1/\alpha$ correction to the pure GB solution~(\ref{eq:largealpha-expand-ngtr2})
is given by the combination of eq.~(\ref{eq:EGBeqnABCF}) expanded in $1/\alpha$
\begin{align}
&\left(\frac{r}{r_0}\right)^\frac{n+2}{2}\left(\frac{f(r)}{n (n+1)}\times (\ref{eq:eqnA}) +\frac{2 }{n (n+1)
   (n+2)}\times (\ref{eq:eqnB}) -\frac{2  f(r)}{n (n+1) (n+2)}\times (\ref{eq:eqnC})-\frac{  f(r)}{n (n+2)}\times (\ref{eq:eqnF}) \right)\nonum
&   \quad =\left(1-\left(\frac{r_0}{r}\right)^{\frac{n-2}{2}}\right) r^2 a_1''(r)+\left(\frac{n}{2}-\left(\frac{r_0}{r}\right)^{\frac{n-2}{2}}\right)   r \, a_1'(r)-\frac{n+4}{2   n^2+2 n} \frac{r^2}{r_0^2}
   +\ord{\alpha^{-1}},
\end{align}
and
\begin{align}
&\frac{r^{n-1}}{r_0^n} \left( -2  f(r) \times  (\ref{eq:eqnA}) -\frac{8  }{n+2} \times (\ref{eq:eqnB})+\frac{8    f(r)}{n+2}\times (\ref{eq:eqnC})+\frac{4  (n+1)  f(r)}{n+2}\times (\ref{eq:eqnF})\right)\nonum
& =\frac{d}{dr} \left[ n (n+1)(2-n) a_1(r)+4 n (n^2-1)  \left( \frac{r}{r_0}\right)^{\frac{n-2}{2}}f_1(r)
 -\frac{2 (n-2)(n+3) }{n+2} \left(\frac{r}{r_0}\right)^\frac{n+2}{2}\right]
 +\ord{\alpha^{-1}},
\end{align}
which lead to eqs.~(\ref{eq:largeA-corr-eq-a1}) and (\ref{eq:largeA-corr-eq-f1-int}), respectively.
The expansion of eq.~(\ref{eq:eqnC}) also leads to eq.~(\ref{eq:largeA-corr-eq-b1}).

\section{An integration in $1/\alpha$ expansion}
Here we study the asymptotic behavior around $x=0$ ($r \gg r_0$) of the integration~(\ref{eq:largeA-def_Fn}), which is rewritten as
\begin{align}
{\sf F}_n(x)& := \int^1_x \frac{y^\frac{n+2}{2-n}-1}{1-y}dy\\
& =  I_n(x) + (n-2)\left[\frac{ x^{4-\frac{4}{n-2}}-1}{4 (n-3)}+\frac{ x^{3-\frac{4}{n-2}}-1}{3 n-10}
+\frac{x^{2-\frac{4}{n-2}}-1}{2  (n-4)}
   +\frac{x^{1-\frac{4}{n-2}}-1}{n-6}+\frac{x^{-\frac{4}{n-2}}-1}{4}\right].\label{eq:def-Fn}
\end{align}
where we defined a finite function $I_n(x)$ for $n>2$ by the following integral
\begin{align}
I_3(x) =0 ,\quad I_n(x) := \int^1_x \frac{1-y^\frac{4(n-3)}{n-2}}{1-y}dy\quad (n>3).
\end{align}
The integration formula
\begin{align}
 \int_0^1 \frac{1-x^{\mu-1}}{1-x} dx = H_{\mu-1} \quad ({\rm Re}(\mu)>0)
 %= \psi(\mu) + \gamma \quad ({\rm Re}(\mu)>0)
\end{align}
guarantees that this function takes a finite value at $x=0$
\begin{align}
 I_n(0) =H_{\frac{4(n-3)}{n-2}},
 %\psi\left(\frac{5n-14}{n-2}\right)+\gamma,
\end{align}
%where $\psi(\mu)$ and $\gamma$ are the digamma function and the Euler constant respectively.
where $H_{\mu}$ is the Harmonic number, whose noninteger values are defined by the digamma function and the Euler constant $ H_{\mu} := \psi(\mu+1)+\gamma$.
For $n=3,4,6$ cases, although the expression~(\ref{eq:def-Fn}) seems singular, one can obtain the regular expression by taking the continuous limit $n \to 3,4,6$, which leads to $\ln x\propto \ln r$ behavior for $r\gg r_0$.
The dominant behavior in ${\sf F}_n(x)$ for $x \ll 1$ becomes
\begin{align}
 {\sf F}_n(x) \simeq \frac{n-2}{4} x^{-\frac{4}{n-2}}\quad (x\ll1),
\end{align}
which gives the asymptotic behavior for $r \gg r_0$
\begin{align}
{\sf F}_n((r_0/r)^\frac{n-2}{2}) \simeq \frac{n-2}{4} \frac{r^2}{r_0^2}.\label{eq:app-Fx-as}
\end{align}

\section{Linear perturbation from GR at small $\alpha_{\rm GB}$}
Here we extend the small $\alpha_{\rm GB}$ analysis of the black string in ref.~\cite{Brihaye:2010me}
 to the arbitrary dimension. Let us consider $\ord{\alpha_{\rm GB}}$ correction to the GR black string solution
\begin{align}
 a = 1 + \frac{\alpha_{\rm GB}}{r_0^2}\, \tilde{a}_1,\quad b = 1-\frac{r_0^n}{r^n} +  \frac{\alpha_{\rm GB}}{r_0^2} \, \tilde{b}_1,\quad f = 1-\frac{r_0^n}{r^n} + \frac{\alpha_{\rm GB}}{r_0^2}\, \tilde{f}_1.
\end{align}
The linear perturbation is solved as
\begin{align}
\tilde{a}_1 =- \frac{n+1}{n+2} \left[ \frac{nr_0^{2n+2}}{r^{2n+2}} \,{}_2 \- F_1\left(1,2+\frac{2}{n},3+\frac{2}{n};\frac{r_0^n}{r^{n}}\right) - 2(n+1) \log \left(1-\frac{r_0^n}{r^{n}}\right) \right],
\end{align}

\begin{align}
& \tilde{b}_1 = -\frac{n (n+1)H_{\frac{n+2}{n}}+n(n^2+n-2)   }{n+2 }\frac{r_0^n}{r^n} 
+\left(\frac{2(n+1)}{n+2}-(n+1)\frac{r_0^n}{r^n}\right)\log\left(1-\frac{r_0^n}{r^n}\right)
   \nonum
      &+(n^2+n-1)\frac{r_0^{2n+2}}{r^{2 n+2}}\left[\frac{2 (n+1)  \,   _2F_1\left(2,\frac{n+2}{n},2+\frac{2}{n};\frac{r_0^n}{r^n}\right)}{n+2}
   -1 \right]
\nonum
   &+ \frac{r_0^{3 n+2}}{r^{3 n+2}}\left[\frac{n (3n+4)(n+1)  \,   _2F_1\left(1,3+\frac{2}{n},4+\frac{2}{n};\frac{r_0^n}{r^n}\right)}{(3n+2)(n+2)} -\frac{\left(4 n^3+15 n^2+8 n-8\right) \,   _2F_1\left(2,2+\frac{2}{n},3+\frac{2}{n};\frac{r_0^n}{r^n}\right)}{2 (n+2)}   \right]
   \nonum
   &+ \frac{r_0^{4 n+2}}{r^{4 n+2}}\left[\frac{2 \left(n^3+3 n^2+n-1\right)   \,_2F_1\left(2,3+\frac{2}{n},4+\frac{2}{n};\frac{r_0^n}{r^n}\right)}{3 n+2}
 +\frac{n^2  (n+1)  \,   _2F_1\left(2,4+\frac{2}{n},5+\frac{2}{n};\frac{r_0^n}{r^n}\right)}{2 (n+2) (2
   n+1)}\right]
\end{align}
and
\begin{align}
\tilde{f}_1 =&-\frac{n r_0^n}{r^n}\left[\frac{(n+1)H_{\frac{n+2}{n}}+n^2-n-4 }{n+2}-\frac{ (n+1)  \log \left(1-\frac{r_0^n}{r^n}\right)}{n+2}\right]\nonum
   &+\frac{r_0^{2n+2}}{r^{2n+2}}\frac{ \left(2  (n+1)(n^2+n-1) \,   _2F_1\left(2,1+\frac{2}{n},2+\frac{2}{n};\frac{r_0^n}{r^n}\right)-n^3-5 n^2-4
   n+2\right)}{n+2}\nonum
   &-\frac{ r_0^{3 n+2}}{ r^{3 n+2}}\left[\frac{\left(4 n^3+15 n^2+8 n-8\right) \,
   _2F_1\left(2,2+\frac{2}{n},3+\frac{2}{n};\frac{r_0^n}{r^n}\right)}{2 (n+2)}-n\right]\nonum
   & +\frac{n+1}{3n+2}\frac{r_0^{4n+2}}{r^{4n+2}}\left[\frac{n  (3n+4)  \,   _2F_1\left(1,3+\frac{2}{n},4+\frac{2}{n};\frac{r_0^n}{r^n}\right)}{n+2}+2 (n^2+2n-1)  \,   _2F_1\left(2,3+\frac{2}{n},4+\frac{2}{n};\frac{r_0^n}{r^n}\right)\right]\nonum
&  + \frac{n^2 (n+1) \,
   _2F_1\left(2,4+\frac{2}{n},5+\frac{2}{n};\frac{r_0^n}{r^n}\right)}{2 (n+2) (2 n+1)}\frac{r_0^{5n+2}}{r^{5n+2}},
\end{align}
where $H_k$ is the Harmonic number and $_2F_1(a,b,c;x)$ is the hypergeometric function. We imposed the boundary condition as
\begin{align}
&  \tilde{a}_1 \to 0 ,\quad   \tilde{b}_1 \to 0 ,\quad   \tilde{f}_1 \to 0,\quad ( r \to \infty),\\
& \tilde{a}_1 \to \ord{1},\quad \tilde{b}_1 \to \ord{r-r_0},\quad \tilde{f}_1 \to \ord{r-r_0},\quad ( r \to r_0).
\end{align}
Up to the linear order, the thermodynamic variables are obtained as
\begin{align}
 \frac{M}{M_{\rm GR}} =1 +\left(\frac{n(n+1)H_{\frac{n+2}{n}}}{n+2}+n(n-1)\right) \frac{\alpha_{\rm GB}}{r_0^2}
\end{align}
\begin{align}
\frac{\tau}{{\tau}_{\rm GR}} = 1 + \left(\frac{n(n+1) H_{\frac{n+2}{n}}}{n+2} - n(n+3)\right) \frac{\alpha_{\rm GB}}{r_0^2}
\end{align}
\begin{align}
\frac{T_{\rm H}}{T_{\rm H,GR}} = 1 - \left(\frac{(n+1)H_{\frac{n+2}{n}}}{n+2} +(n+2)(n-1)\right) \frac{\alpha_{\rm GB}}{r_0^2}
\end{align}
\begin{align}
\frac{S}{S_{\rm GR}} = 1  + \left(\frac{(n+1)^2 H_{\frac{n+2}{n}}}{n+2} + 2n(n+1)\right) \frac{\alpha_{\rm GB}}{r_0^2}
\end{align}
where
\begin{align}
M_{\rm GR} := \frac{(n+1)\Omega_{n+1} L r_0^n}{16 \pi G},\quad \tau_{\rm GR} := \frac{\Omega_{n+1}  r_0^n}{16 \pi G},\quad
T_{\rm H,GR} := \frac{n}{4\pi r_0},\quad S_{\rm GR} := \frac{\Omega_{n+1} L r_0^{n+1}}{4 G}\label{eq:thermovars-GR}
\end{align}
The relative tension is given by
\begin{align}
  N = \fr{n+1} \left(1 - \frac{2 n(n+1)\alpha_{\rm GB}}{r_0^2}\right).
  \label{eq:N-lin}
\end{align}
All these are consistent with the result in ref.~\cite{Brihaye:2010me} for $n=1,\dots,4$.

\end{document}